\documentclass[fleqn,usenatbib]{mnras}

\usepackage{newtxtext,newtxmath}

\usepackage[T1]{fontenc}
\DeclareRobustCommand{\VAN}[3]{#2}
\let\VANthebibliography\thebibliography
\def\thebibliography{\DeclareRobustCommand{\VAN}[3]{##3}\VANthebibliography}

\usepackage{graphicx}	% Including figure files
\usepackage{amsmath}	% Advanced maths commands
\usepackage{xcolor}
\usepackage{physics}
\usepackage{float}
\usepackage{caption}
\usepackage{subfigure}
\usepackage{bm}
\usepackage{gensymb}
\usepackage{booktabs}
\newcommand{\D}{\mathrm{d}}
\newcommand\bv{Brunt-Väisälä frequency }

\newcommand\bvns{Brunt-Väisälä frequency}
\newcommand\omegabv{\omega_\mathrm{BV}^2}
\newcommand\ye{Y_{\!\mathrm{e}}}
\newcommand{\ud}{\mathrm{d}}

\title[Convection and Core $g$-mode in PCSs]{Convection and the Core $g$-mode in Proto-Compact Stars -- A detailed analysis}

\author[Jakobus et al.]{
Pia Jakobus,$^{1,2}$\thanks{E-mail: pia.jakobus@uni-hamburg.de}
Bernhard M\"uller,$^{2,3}$\thanks{E-mail: bernhard.mueller@monash.edu}
and Alexander Heger$^{2,3}$\thanks{E-mail: alexander.heger@monash.edu}
\\
$^{1}$University of Hamburg, Hamburger Sternwarte, Gojenbergsweg 112, 21029, Hamburg, Germany\\
$^{2}$School of Physics and Astronomy, Monash University, VIC 3800, Australia\\
$^{3}$ ARC Centre of Excellence for Gravitational Wave Discovery -- OzGrav
}

\date{Accepted XXX. Received YYY; in original form ZZZ}

\pubyear{2015}

\begin{document}
\label{firstpage}
\pagerange{\pageref{firstpage}--\pageref{lastpage}}
\maketitle

\begin{abstract}
We present a detailed analysis of the dynamics of proto-compact star (PCS) convection and the core ${}^2\!g_1$-mode in core-collapse supernovae based on general relativistic 2D and 3D neutrino hydrodynamics simulations. Based on 2D simulations, we derive a mode relation for the core $g$-mode frequency in terms of PCS and equation of state parameters, and discuss its limits of accuracy. This relation may prove useful for parameter inference from future supernova gravitational wave (GW) signals if the core $g$-mode or an emission gap at the avoided crossing with the fundamental mode can be detected. The current 3D simulation does not show GW emission from the core $g$-mode due to less power in high-frequency convective motions to excite the mode, however. Analysing the dynamics of PCS convection in 3D, we find that simple scaling laws for convective velocity from mixing-length theory (MLT) do not apply. Energy and lepton number transport are instead governed by a more complex balance between neutrino fluxes and turbulent fluxes that results in roughly uniform rates of change of entropy and lepton number in the PCS convection zone. Electron fraction and enthalpy contrasts in PCS convection are poorly captured by the MLT gradient approximation. We find distinctly different spectra for the turbulent kinetic energy and turbulent fluctuations in the electron fraction, which scale approximately as $l^{-1}$ without a downturn at low $l$. We suggest that the different turbulence spectrum of the electron fraction is naturally expected for a passive scalar quantity.
\end{abstract}

\begin{keywords}
gravitational waves -- transients: supernovae -- convection
\end{keywords}

\section{Introduction}
Ordinary core-collapse supernovae (CCSNe) with explosion energies of $\mathord{\sim} 10^{51}\,\mathrm{erg}$ explode via the delayed neutrino-driven mechanism~\citep{Colgate_1961}. In the neutrino-driven paradigm the shock wave formed during core bounce stalls at a radius of about $100\,\mathrm{km}- 200\,\mathrm{km}$, and gets revitalised by partial re-absorption of neutrinos emitted from the proto-compact star (PCS). This leads to the expulsion of outer layers of the star~\citep{Janka_2016,mueller_2020,Burrows_review}. 

Multi-dimensional effects are critical for the success of neutrino-driven explosions and highly relevant to observations as they imprint characteristic signatures on the neutrino and gravitational wave signals of the supernova core \citep{mueller_2019b,abdikamalov_22}.
Neutrino heating drives convective overturn in the gain region behind the supernova shock \citep{Herant_1992,Herant_1994,Burrows_1995,Janka_1996}. In addition, large-scale shock oscillations, i.e., the standing accretion shock instability (SASI) can develop due to a vortical-acoustic feedback cycle \citep{blondin_2003,Guilet_2012}.
Convection develops inside the newly formed PCS after the collapse of the iron core between radii of $10\,\mathrm{km} \mathord{\lesssim} r \mathord{\lesssim} 25\,\mathrm{km}$ due to energy and lepton-number losses at the PCS surface. 
Early theoretical considerations suggested that convective instabilities beneath the neutrinosphere play a vital role in CCSN dynamics~\citep{Epstein_1979,Bruenn_1979,Livio_1980}. 
As PCS convection significantly affects neutrino luminosities and mean energies, it has a potentially relevant impact on the neutrino heating conditions and hence the conditions for explosion~\citep{Keil_1996}. However, later studies by \cite{Dessart_2006,Buras_2006a} have shifted the focus away from the dynamical role of PCS convection in the explosion mechanism.
As PCS convection efficiently transports energy and lepton number out of the PCS core, it does, however, affect the observable neutrino signal and the PCS cooling evolution~\citep{roberts_12,Mirizzi_2016,Pascal_2022}.

PCS convection also contributes to gravitational-wave (GW) emission
from CCSNe. Along with aspherical fluid motions in the gain region, it excites oscillation of the quadrupolar $f$/$g$ surface oscillation mode that constitutes the most prominent feature in the supernova GW signal. It may even be a GW source in its own right \citep{Murphy_2009,mueller_2013,andresen_2017,Morozova_2018,Radice_2019,Mezzacappa_2020,Vartanyan_2023}. More recently, the ${}^2\!g_1$-mode in the PCS core has received
attention. Its primary forcing mechanism is PCS convection. This \textit{core} $g$-mode has been found in 2D simulations~\citep{pablo_2013,Kawahara_2018,torres_2019b,Jakobus_2023} and recently also in 3D~\citep{Vartanyan_2023}, and is of interest as a more direct probe of the high-density equation of state than the surface $f$/$g$-mode. As demonstrated by \citet{Jakobus_2023}, the ${}^2\!g_1$-mode is sensitive to the speed of sound at around twice saturation density. The conditions for the excitation of the core ${}^2\!g_1$-mode are, however, still less clear than for the surface $f$/$g$-mode. Moreover, the relation of the 
${}^2\!g_1$-mode to the physical parameters of the PCS are not yet as well understood as for the surface $f$/$g$-mode. For the surface $f$/$g$-mode, various physically motivated fitting formulas have been proposed to relate the mode frequency to the PCS mass, radius, and surface temperature 
\citep{mueller_2013,Morozova_2018,torres_2019b,sotani_21,Mori_2023}.
As of yet, little effort has been made to find analytically motivated relations for the ${}^2\!g_1$-mode. 
In this study, we aim for a simple ``semi-analytic'' estimate of the \textit{core} $g$-mode, analogously to the analysis done for the surface quadrupolar $g$-mode in \citet{mueller_2013}. This could potentially enable parameter estimation of PNS core properties and equation of state (EoS) parameters based on the prospective GW signal from the quadrupolar core $g$-mode.

Related to the GW emission and mode excitation by PCS
convection, there are also open questions about the dynamics of the convective flow. In the case of convection in the gain region,
the saturation of turbulent kinetic energy \citep{mueller_2020}
and the relation between turbulent kinetic energy and the emission of GWs via excitation of the surface $f/g$-mode
\citep{mueller_17gw,powell_2019,Radice_2019} have been investigated extensively. 
The impact of dimensionality on convection in the gain region has also been thoroughly investigated
\citep[e.g.,][]{Hanke_2012,Dolence_2013,Couch_2013}, identifying critical differences in the dynamics of neutrino-driven convection in two (2D) and three dimensions (3D)
due to the inverse turbulent cascade in 2D \citep{Kraichnan_1967}.
Empirically, it has been found that GW emission due to
the excitation of the surface $f/g$-mode by convection in the gain region is weaker in {3D than in 2D, which has been attributed to smaller terminal downflow velocities and less high-frequency variability in the downflows \citep{andresen_2017}.  
By and large, these issues have been less well explored for PCS convection, although the phenomenology of PCS convection across the progenitor mass range has recently been studied by \citet{nagakura_2020}.

One of the more puzzling aspects of PCS convection
is the lepton-number emission self-sustained asymmetry (LESA) that has been found in many 3D simulations \citep{Tamborra_2014,Vartanyan_2019,OConnor_2018,powell_2019,Janka_2016,glas_19,Vartanyan_2019b,nagakura_2020} and still remains incompletely understood. 
In LESA, the neutrino emission develops a dipolar asymmetry several hundreds of milliseconds after core-bounce. The hemispheric luminosity difference of $\nu_e$ and $\overline{\nu}_e$ neutrinos can amount to tens of percent. Different from the initial proposition of a feedback mechanism between accretion and PCS convection \citep{Tamborra_2014}, the mechanism responsible for LESA may operate entirely within the PCS
convective region~\citep{glas_19}. 
\citet{Janka_2016} identified complex flow dynamics in LESA
due to the sensitivity of the
Ledoux criterion to the (negative) radial  electron fraction  gradient $\D\ye/\D r$, which can become stabilising depending on
the thermodynamic derivative$\left(\partial\rho/\partial\ye\right)_{P,s}$.
This can inhibit convection in certain directions and, in turn, lead to less effective transport of the lepton number upwards from the lepton-rich inner core. In conjunction with neutrino diffusion, a positive feedback cycle can be envisaged that gives rise to a sustained global asymmetry in the election fraction.
To explain the low-mode nature of LESA,
the study by \citet{glas_19} drew an analogy between the excitation of harmonics of different orders in thermally unstable spherical shells according to Chandrasekhar’s linear theory of thermal instability in spherical shells \citep{Chandrasekhar_1961}. Building further on this, \citet{powell_2019} suggested LESA not to be a new instability but instead a manifestation of convection, but with a non-Kolmogorov velocity spectrum due to the presence of a stabilising $Y_\mathrm{e}$-gradient in the middle of the convection zone.
In this paper, we further address these current topics in PCS convection. Following our recent study of GW emission from the core $g$-mode \citep{Jakobus_2023}, we conduct 2D and 3D simulations with the relativistic neutrino hydrodynamics code \textsc{CoCoNuT-FMT} \citep{mueller_2010,mueller_2015} for
massive zero-metallicity progenitor stars
of $35 \mathrm{M}_\odot$ and $85 \mathrm{M}_\odot$ with strong PCS convection.
Based on these simulations, we shall investigate the following questions:
\begin{enumerate}
\item
What determines the trajectory of the core $g$-mode frequency, and can this frequency be related to PCS properties? 
    \item 
    How does the amplitude of the core $g$-mode signal depend on dimensionality?
    \item 
    What determines the convective velocities inside the PCS convection zone and how well are they described by mixing-length theory? 
    \item 
    What determines the spectrum of $Y_
    \mathrm{e}$-fluctuations in the PCS convection zone (and possibly the dominance of the dipole in LESA)? 
\end{enumerate}
The structure of this paper is as follows: In Section~\ref{sec:methods}, we give an overview of the numerical methods, progenitor setup and the equation of state. 
In Section~\ref{sec:GW} we analyse the GW signals of our models. We discuss the GW signals of our 2D and 3D simulations, and then develop approximations for the frequency trajectory of the core $g$-mode.  In Section~\ref{sec:II}, we investigate the properties of the PCS convection zone. We briefly discuss basic explosion properties, analyse the structure of turbulent motion, present turbulent kinetic energies in the PCS convection zone for both 2D and 3D,  temporal fluctuations in PCS convection, and the efficiency of turbulent convection. We then discuss to what extent the turbulent convective velocities are captured by balancing principles and mixing-length theory (MLT). Finally, we address the spectrum of 
velocity and passive scalars in the PCS convection zones.
A summary of our findings and conclusions are presented in Section~\ref{sec:concl}.

\section{Methods}\label{sec:methods}
We use the neutrino-hydrodynamic code CoCoNuT-FMT for our 2D and 3D CCSN simulations. The equations of hydrodynamics are solved with a general relativistic finite-volume based solver~\citep{mueller_2010}. For the neutrino-transport, we use the fast multigroup transport method of \citet{mueller_15b}. For computational efficiency, the innermost zones are treated in spherical symmetry. To capture the entire PCS convection zone in 2D or 3D, we decrease the inner boundary of the spherical core to $3.1\,\mathrm{km}$ before PCS convection develops.

At high densities, we consider two different nuclear equations of state. In one set, we employ the CMF equation of state with a first-order nuclear liquid-vapour phase transition at densities~$\mathord{\sim}\rho_0$, a second (weak) first-order phase transition due to chiral symmetry restoration at {around four times nuclear saturation density $\rho_0$ ($\rho_0\approx 2.6\times 10^{14}\mathrm{gcm}^{-3}$)} with a critical endpoint at $T_\mathrm{CeP}= 15\,\mathrm{MeV}$, and a smooth transition to quark matter at higher densities~\citep{motornenko_2020}. 
We also consider one model with the SFHx EoS \citep{Steiner_13}. The high-density EoS is denoted by a model suffix \texttt{\_CMF}
or \texttt{\_SFHx}, respectively. At low densities, the equation of state includes 20 species of non-interacting nucleons and nuclei, electrons, positrons, and photons. Nuclear reactions are treated by the flashing scheme of
\citet{rampp_2000}. At a temperature above $0.5\,\mathrm{MeV}$, nuclear statistical equilibrium is assumed.

As in \citet{Jakobus_2023}, we use two zero-metallicity progenitors of $35\,\mathrm{M}_\odot$ and $85\,\mathrm{M}_\odot$, named \texttt{z35} and \texttt{z85}, \citep{heger_2010}, and additionally simulate a zero-metallicity $85\,\mathrm{M}_\odot$ progenitor in 3D (utilising the CMF EoS); all progenitors are calculated with the stellar evolution code \textsc{Kepler}~\citep{weaver_1978}.  

Gravitational wave signals are computed with the time-integrated quadrupole formula with general relativistic corrections \citep{mueller_2013}.

\section{Gravitational Wave Emission}\label{sec:GW}
In this section, we compare GW spectrograms of the 2D and 3D simulations of the $85\,\mathrm{M}_\odot$ progenitor with the CMF EoS. Subsequently, we develop approximations for the core $g$-mode frequency based on the GW spectrograms of the $85\,\mathrm{M}_\odot$ and $35\,\mathrm{M}_\odot$ progenitors simulated in 2D in \citet{Jakobus_2023}, with the CMF and SFHx EoS \citep{motornenko_2020,Steiner_13}. 

\subsection{Comparison of Spectrograms in 2D and 3D}\label{gw_discussion}
\begin{figure*}
    \centering
    \includegraphics[width=0.8\linewidth]{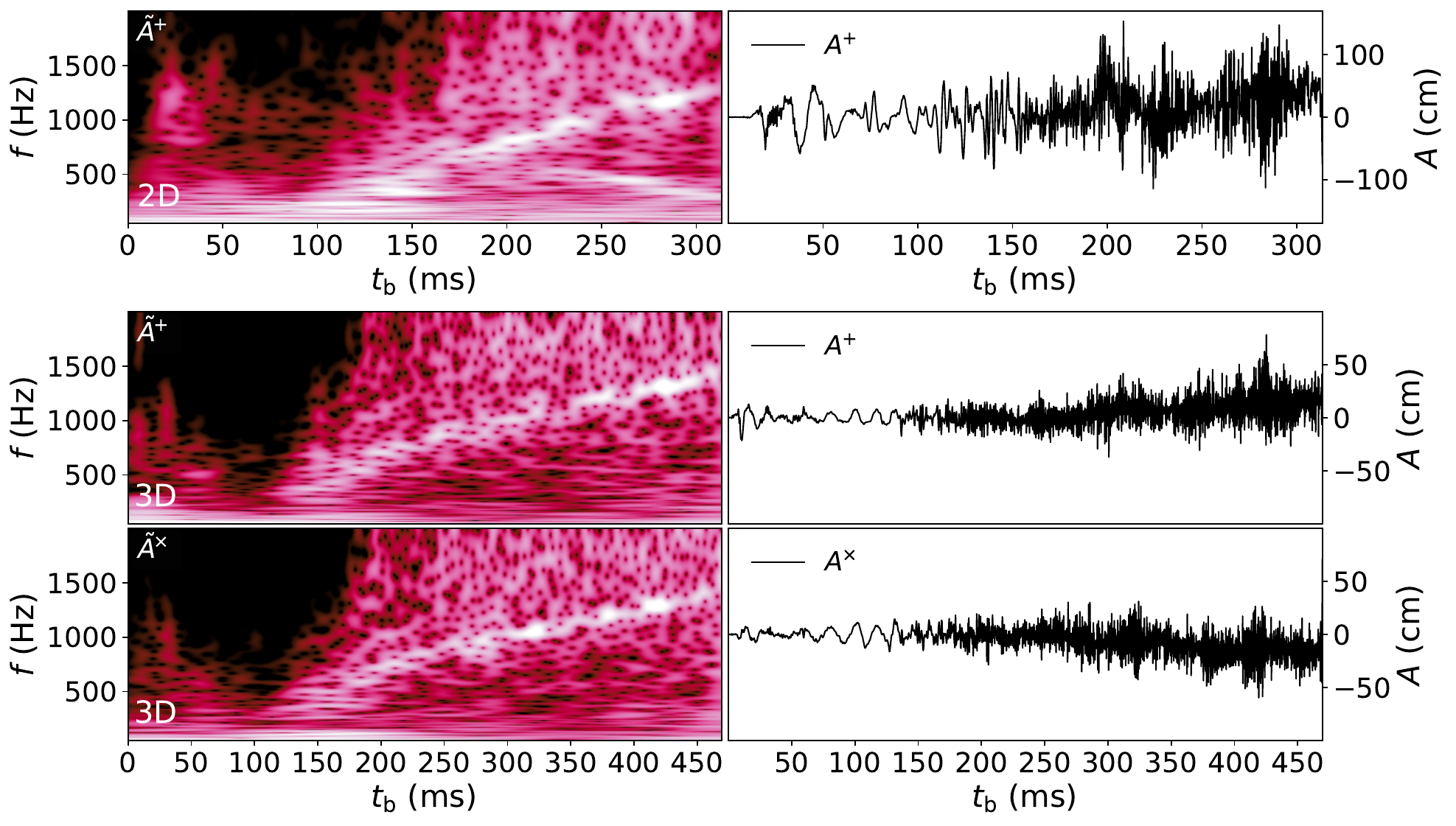}
    \caption{GW  spectrograms and amplitudes for models \texttt{z85} for 2D ({upper panel}) and 3D ({lower two panels}). Plus and cross amplitudes are evaluated for an observer in the equatorial plane. The CMF-EoS is employed in both runs. Note, that the 2D run exhibits a higher maximum scale for the amplitude $|\tilde{A}|^+$ (not shown) as GW amplitudes are generally increased in 2D compared to 3D. 
 The 2D model exhibits the distinct second frequency band from the ${}^2\!g_1$-mode, which branches of the dominant band after a few hundred milliseconds. The core $g$-mode signal is absent in 3D.}
    \label{fig:gw_spectro_3d2d}
\end{figure*}
Figure~\ref{fig:gw_spectro_3d2d} shows GW spectrograms obtained with the \href{https://docs.scipy.org/doc/scipy/reference/generated/scipy.signal.morlet.html}{Morlet wavelet transform} \citep{Morlet_1982}, together with the signal amplitude. The GW signal for the 2D run of model \texttt{z85\_CMF}  is shown in the {upper panel} (plus polarisation only), and results (for both the plus and cross polarisation) from the corresponding 3D run are shown {in the second and third panel, respectively}. The signals are shown 
for an observer at the equator at $(\theta,\phi)=(90\degree,90\degree)$. 

The 2D run exhibits significantly larger GW amplitudes with $A^{\!+} >150\,\mathrm{cm}$, compared to $A^{\,+} \lesssim 80\,\mathrm{cm}$ in the 3D model. This is a known feature and has been explained by smaller convective structures, slower overturn velocities, and less impulsive forcing on the PCS surface in 3D \citep{Mueller_1997,mueller_2012,andresen_2017}. 
Nonetheless, the amplitudes of the 3D model are very high for both the plus and cross components (up to $A^{\!+}\mathord{\sim} 80\,\mathrm{cm}$). This can be compared to the results of \citet{powell_21}, who simulated the same progenitor with the \texttt{SFHx}-EoS and found smaller amplitudes of the order $A^{\!+}\mathord{\sim}25\,\mathrm{cm}$.  
In the initial phase, the spectrograms of the 2D and 3D simulations are very similar. Both models display low-frequency GW emission at around $100\,\mathrm{Hz}$. This is due primarily to the presence of prompt convection and early SASI activity \citep{Marek_2009, Murphy_2009, Yakunin_2010, mueller_2013}. The 2D model shows stronger low-frequency activity than the 3D model throughout the simulation.
Later on, as emission moves to higher frequencies, the dominant $f/g$-mode emission band emerges in both models, starting at around $\mathord{\sim} 300\,\mathrm{Hz}$ and reaching frequencies above $1400 \,\mathrm{Hz}$ in 3D. In 2D, the evolution is cut off at $\mathord{\sim}1250\,\mathrm{Hz}$ because a black hole forms early on. We note that the $f/g$-mode frequency increases slightly faster in the 2D simulation.
One key difference between the spectrograms of both models is that the core $g$-mode is absent in 3D. This particular ${}^2\!g_1$ mode is typically located in deep regions of the PCS at radii $5\,\mathrm{km}\lesssim R\lesssim 15\,\mathrm{km}$. Previous 2D studies have identified this mode \citep{pablo_2013, Kawahara_2018, torres_2019, Jakobus_2023} and linked to the speed of sound \citep{Jakobus_2023}; more recently, it has also been identified in 3D \citep{Vartanyan_2023}. GW emission from the core $g$-mode is thus a less robust feature than the signal from the $f/g$-mode. This does not imply that the core $g$-mode is a 2D artefact. Its absence in the current simulation may be compounded by stochastic model variations. More high-resolution 3D models are required to decide whether and when it appears and whether it is potentially detectable.

\subsection{Core $g$-mode frequency relation}\label{sec:fit}
If the GW signal from the core $g$-mode is present and detectable under certain circumstances, it is important to connect the mode frequency to physical parameters of the PCS and the EoS. Even if the core $g$-mode is not present or only visible in the spectrogram for a shorter time, the avoided crossing\footnote{Avoided crossing of frequencies typically occurs in linear oscillation analysis when different types of modes have similar frequencies \citep{Torres_2018}. In the regions of avoided crossing, the eigenfunctions of the two modes take on a mixed character \citep{Aizenman_1977,Stergioulas_2003}.} with the surface $f/g$-mode may be a more robust feature \citep{Morozova_2018,Sotani_2020,Vartanyan_2023} that can potentially be used for inferring PCS or EoS parameters.
{The modes identified in this work are more closely related to interface (or surface) type g-modes, rather than the typical gravity modes that emerge from continuously varying composition or entropy gradients with depth}.
In this section, we use the three 2D models \texttt{z35:CMF}, \texttt{z85:SFHx}, and \texttt{z85:CMF} from \citet{Jakobus_2023} to  develop and assess analytic approximations for the core $g$-mode frequency in terms of EoS and PCS parameters\footnote{No signal was observed in the 2D run \texttt{z85:SFHx}}.

For approximating the mode frequency $f_{^2g_1}$ analytically, we note that the frequency of low-order $g$-modes living in a convectively stable region often scales well
with the maximum of the \bv, i.e., $f_{^2g_1}\propto \omega_\mathrm{BV}$, with a proportionality factor of order unity. The maximum of the \bv is located at the edge of the PCS core and associated with a recognisable entropy step between the core and the PCS convection zone. For the first few hundred milliseconds after bounce, the entropy step established by the early propagation of the shock will not change substantially due to efficient neutrino trapping, and we can therefore assume to
first order that the location of the step $M_\mathrm{fix}$ in mass coordinate remains constant (though perhaps somewhat EoS-dependent), and that the
entropy and electron fraction gradients at this point also remain roughly constant.
This approximation will, of course, eventually break down at later post-bounce times.

The task is then to approximate the maximum of $\omega_\mathrm{BV}$, which can be written in general relativity as \citep{Jakobus_2023} {by expanding the radial entropy and electron fraction derivatives\footnote{giving rise to fixed thermodynamic quantities, shown as subscripts} from the standard
Ledoux criterion in terms of thermodynamic parameters \citep{Ledoux_1947}, i.e., the entropy and electron fraction gradient and thermodynamic derivatives of the pressure},
\begin{align}\label{eq:bv1}
    \omega_\mathrm{BV}^2 &= \underbrace{\dv{\alpha}{r}\frac{\alpha}{h \phi^4}}_{\mathrm{``effective\,rel.\,acc.''}}\underbrace{\frac{1}{\rho c_\mathrm{s}^2}\left\{\left(\pdv{P}{s}\right)_{\rho,Y_\mathrm{e}}\dv{s}{r} + \left(\pdv{P}{Y_\mathrm{e}}\right)_{\rho,s}\dv{Y_\mathrm{e}}{r}\right\}}_{\mathrm{``EoS\,\,factors''}},
\end{align}
for a conformally flat metric.
Here $r$ is the radial coordinate, $\alpha$ is the lapse function, $\phi$ is the conformal factor, $c_\mathrm{s}$ is the sound speed,  $P$ is the pressure, $s$ is the specific entropy,
$Y_\mathrm{e}$ is the electron fraction, and
$h$ is the relativistic enthalpy, which is defined in terms of $P$, the density $\rho$ and the specific internal energy $\epsilon$ as $h=1+\epsilon+P/\rho$.
The right-hand side of Equation~(\ref{eq:bv1}) can conveniently be split into an effective
relativistic acceleration term (as a generalisation of the Newtonian gravitational acceleration) and factors
that depend on the EoS and radial derivatives of thermodynamics quantities (``EoS factors'').

To this end, we seek suitable approximations for the metric functions $\phi$ and $\alpha$, and the derivative $\D\alpha/\D r$, and for the radial gradient of the entropy $\D s/\D r$ at the edge of the core, which determines the stiffness of the convectively stable region. This enables us to formulate an EoS-dependent mode relation, which could constrain the EoS at high densities via the thermodynamic terms that appear in the \bvns. Throughout the following calculations, we will use geometric units for metric quantities.

\subsubsection{Metric terms}
To approximate the metric function $\alpha$ in the ``effective relativistic acceleration'' $c_\mathrm{gr}$ (see prefactor in Eq.~\ref{eq:bv1}),
we work in the Newtonian limit (where $\alpha\approx1+\Phi$ in terms of the Newtonian potential $\Phi$)
and include contributions to the potential from the mass $M_\mathrm{mode}$ inside the boundary between the core and the PCS convection zone and the remaining neutron star mass outside.
We show the detailed calculation for the approximate lapse function in the Appendix \ref{appendix:alpha} and only quote the result,
\begin{align}
    \alpha\approx 1 - G\left(\frac{M_\mathrm{mode}}{r_\mathrm{mode}} - \frac{M_\mathrm{NS} - M_\mathrm{mode}}{\langle r\rangle}\right) + \mathcal{O}(r^{-2}),
\end{align}
where $r_\mathrm{mode}$ is the radius of where the core $g$-mode is located, $M_\mathrm{NS}$ is the baryonic mass of the PCS. The coefficient
$c_\mathrm{eff}$ is set to $c_\mathrm{eff} = 0.5$.
Finally, $\langle r\rangle$ is an ``effective'' radius for the PCS material outside the core,
\begin{align}
    \langle r\rangle \equiv c_\mathrm{eff} (R_\mathrm{NS} + R_\mathrm{mode}).\label{eq:ceff}
\end{align}
Building on this approximation for $\alpha$, we proceed to estimate the entire effective relativistic acceleration term in Equation~(\ref{eq:bv1}). It is useful to factor out the term $r^{-2}$, so that we can approximate the more slowly varying quantity,
\begin{align}
c_\mathrm{gr}~\equiv~\left[{\D\alpha}/{\D r}\!\cdot r^2 \!{\alpha}/{(h\phi^4)}\right]^{1/2}.
\end{align}
Hence, $\omegabv$ becomes,
\begin{equation}
    \omega_\mathrm{BV}
    =c_\mathrm{gr}
    \sqrt{\frac{1}{r^2\rho c_\mathrm{s}^2}\left\{\left(\pdv{P}{s}\right)_{\rho,Y_\mathrm{e}}\dv{s}{r} + \left(\pdv{P}{Y_\mathrm{e}}\right)_{\rho,s}\dv{Y_\mathrm{e}}{r}\right\}},
\end{equation}
Separating the factor $r^{-2}$ has the added benefit that we can later absorb the term $r^2\rho$ inside the square root by rewriting the radial thermodynamic derivatives as derivatives with respect to enclosed mass $m$.

To find an expression for $c_\mathrm{gr}$, we note
that the lapse function $\alpha$ and conformal factor $\phi$ within the PCS approximately conform to the relation $\alpha\phi^2~\approx~1$~\citep{mueller_2013}. 
Since the thermodynamic structure is related to the gravitational potential in hydrostatic equilibrium, we can also empirically relate the enthalpy $h$ to the conformal factor $\phi$ as $h \approx \phi$.  
In addition, we use the approximation $\D\alpha/\D r \approx G m h / (r \phi^2)^2$, where $r \phi^2$ is the circumferential radius. The factor $h$ provides
a crude correction factor for converting $m$ into the gravitational mass of the PCS core.

Figure~\ref{fig:gr_terms} shows our fits for the individual terms in the relativistic prefactor $c_\mathrm{gr}$ for the 2D model \texttt{z85:CMF}. 
The shaded regions depict fit values for different fixed mass coordinates $M_\mathrm{fix}$. The dashed red line represents the exact value at the peak of the \bvns; the dashed black line shows our analytic estimate evaluated at the mass coordinate corresponding to the peak of the \bvns. Note that the error can not be more accurate than the approximation evaluated at the peak of the \bvns. The two error sources are thus the approximations themselves, and the assumption that the mass coordinate, $M_\mathrm{fix}$, of the entropy step is approximately constant.  
The approximation tracks the lapse function $\sqrt{\alpha}$ with a slight offset, but the overall trend aligns well. 
Early on, the approximation to the derivative $\left(\D\alpha/\D r\right)^{1/2}$ matches very well, whereas it deviates by $\mathord{\sim} 10\%$ at later times. 
At early times, $\D\alpha/\D r$ is quite sensitive to the choice of the fixed mass shell; however, later on, the error arises from the approximation of the derivative of the metric function $\D\alpha/\D r$, and not from the choice of the fixed mass shell (see red line versus black line). Nevertheless, it is clear that a choice of $M_\mathrm{fix}\mathord{\sim} 0.7,\mathrm{M}_\odot$ provides the best fit. 
Our approximation of the enthalpy $\sqrt{h}$ is very accurate, with errors below $\mathord{\sim}3\%$. 
The conformal factor $\phi$ has an acceptable error of~$\mathord{\sim} 6\%$. 
\begin{figure*}
	\centering	\includegraphics[width=0.7\linewidth]{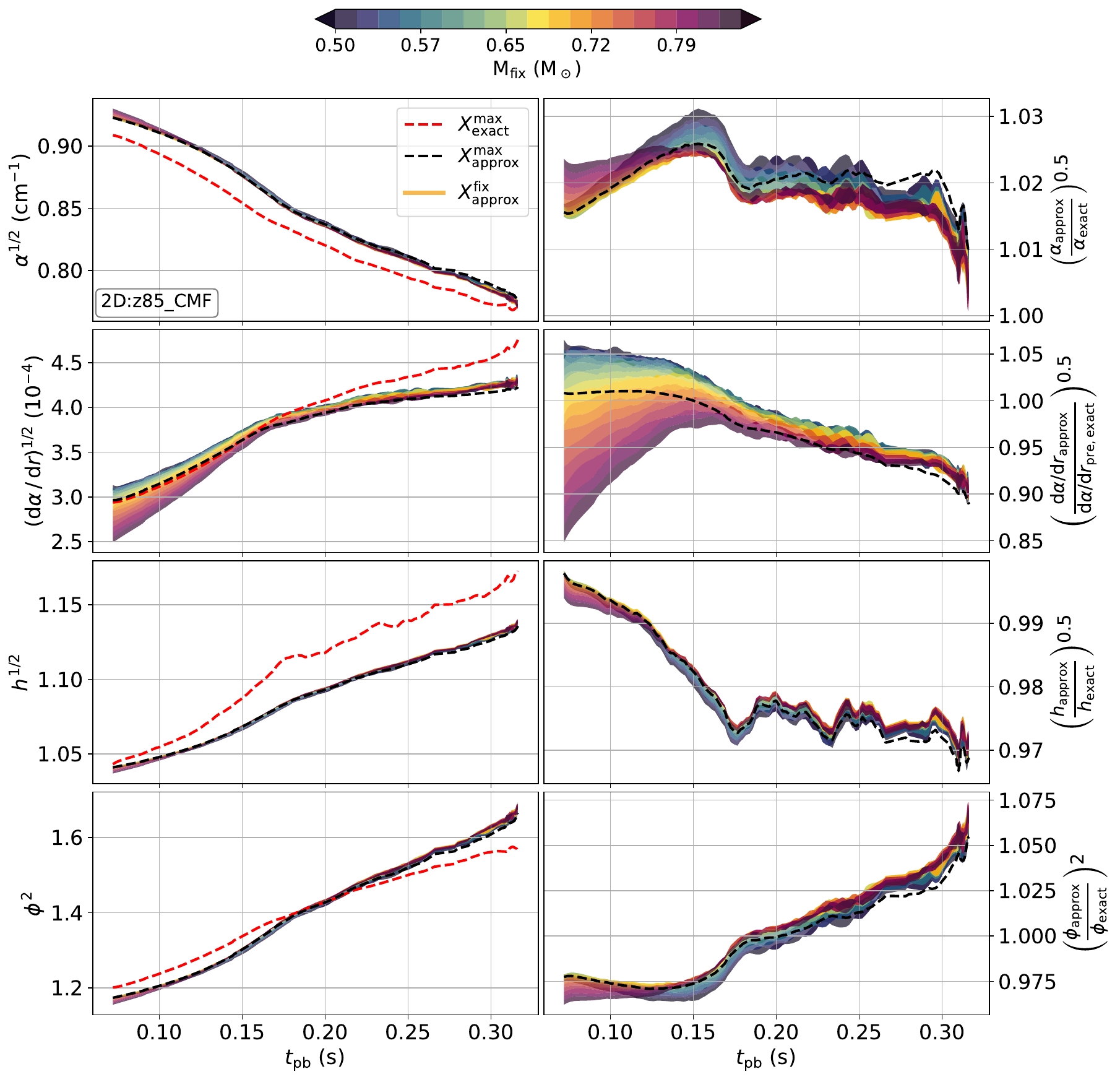}
\caption{Left panel: Approximations for the lapse function $\alpha$, its derivative $\D\alpha/\D r$, conformal factor $\phi$, and relativistic enthalpy $h$ for the 2D model \texttt{z85:CMF}. Right panel:{Relative ratio} of the respective approximations. Red dashed lines show the exact values, black dashed lines show the approximations at the maximum of the Brunt-V\"ais\"al\"a frequency, and coloured lines show the mass coordinates of the fixed-mass shell approximation.
The orange line ($X^\mathrm{fix}_\mathrm{approx}$) corresponds to our default choice for $M_\mathrm{fix}$.
}\label{fig:gr_terms}
\end{figure*}

To conclude, the {relative ratio} at late times in Fig.~\ref{fig:gr_terms} is primarily caused by the term $\left(\D\alpha/\D r\right)^{1/2}$. This error, in turn, is mostly due to the choice of the mass coordinate $M_\mathrm{mode}$ that appears in the approximation for the derivative of the lapse function $\alpha$. Approximating the derivative analytically for any given mass coordinate introduces a smaller error.
We also note that, during the signal interval, the terms appearing in the effective relativistic acceleration are only slightly sensitive to the fixed mass shell.

\subsubsection{EoS and PCS structure terms}
We next turn our attention to the ``EoS factor'' appearing in the \bv in Equqation~(\ref{eq:bv1}). As a first approximation, we neglect the term $\left({\partial P}/{\partial Y_\mathrm{e}}\right)_{\rho,s} \D Y_\mathrm{e}/{\D r}$ as it is about a factor~$\mathord{\sim} 12$ smaller than the first term $\left({\partial P}/{\partial s}\right)_{\rho,Y_\mathrm{e}}{\D s}/{\D r}$ during our the time interval of interest. We define the EoS parameter $c_\mathrm{eos}$ as
\begin{align}
\label{eq:c_eos}
    c_\mathrm{eos} \equiv 
    \left(\pdv{P}{s}\right)^{1/2}_{\rho,Y_\mathrm{e}}/c_\mathrm{s},
\end{align}
so that $\omegabv$ becomes,
\begin{equation}
\label{eq:ombv_rewritten}
    \omega_\mathrm{BV}
    \approx c_\mathrm{gr}
    c_\mathrm{eos}
    \sqrt{\frac{1}{r^2\rho} \dv{s}{r}}
    =  c_\mathrm{gr}
    c_\mathrm{eos}
    \sqrt{\frac{4 \pi r^2\rho}{r^2\rho} \dv{s}{m}}
    = c_\mathrm{gr}
    c_\mathrm{eos}
    \sqrt{4 \pi \dv{s}{m}}.
\end{equation}
The radial profiles of the angle-averaged entropy do not vary substantially for
the first few hundred milliseconds after bounce.
Furthermore, the homologous core mass at bounce and the entropy profiles at the edge of the PCS core are not strongly sensitive to progenitor variations. The sensitivity to progenitor
variations may actually be overaccentuated by
the use of the deleptonisation scheme of
\citet{Liebendoerfer_05b} as an approximation during the collapse phase. This means that we can treat the entropy gradient in 
Equation~(\ref{eq:ombv_rewritten}) as only (weakly) dependent on the EoS, and relatively universal across progenitors {\citep{phdthesis_marik07}}. 
Based on our simulations, we choose a value of $\D s/\D m\approx 11.93 k_\mathrm{B} / \mathrm{M}_\odot$ for the entropy gradient.

To demonstrate our approximation for
$c_\mathrm{eos}$, we show the factors
$c_\mathrm{s}$ and $(\partial P/\partial s)_{\rho,Y_\mathrm{e}}^{1/2}$
in Equation~(\ref{eq:c_eos}) at fixed mass coordinates in Figure~\ref{fig:eos_terms} and compare them to their actual values at the peak of the \bvns. This is similar to our verification of the approximations for the metric terms in Figure~\ref{fig:gr_terms}. As is evident from the relative error in the right column, the best fit for the speed of sound in Panel~1 is again obtained approximately for $M_\mathrm{fix}\approx 0.7\,\mathrm{M}_\odot$, with errors of $\lesssim 10\%$. 
For both quantities, the error increases (see right column), reaches a maximum, and then decreases again: $M_\mathrm{fix}$ is initially chosen at a slightly larger mass coordinate than the core $g$-mode; as the mass coordinate of the peak of the \bv wanders inside and closer to $M_\mathrm{fix}$, the error becomes smaller, and then increases again as the peak of the \bv moves to mass coordinates smaller than $M_\mathrm{fix}$. This behaviour is even better understood from Figure~\ref{fig:moving_gmode}, which shows the angle-averaged sound speed and density profiles, for model \texttt{z85:CMF}. Vertical lines in the top panel denote the mass coordinate of the maximum \bv $\omega_\mathrm{BV}^{2,\mathrm{max}}$ as a function of time; the dashed vertical line in the second row denotes the fixed mass shell coordinate $M_\mathrm{fix}=0.72\,\mathrm{M}_\odot$. The mismatch between $M_\mathrm{fix}$ and $M_\mathrm{\omega_\mathrm{BV}^\mathrm{2,max}}$ is reflected in an underestimation of the sound speed as relevant to the $g$-mode at later times. 
\begin{figure*}
	\centering	\includegraphics[width=0.8\linewidth]{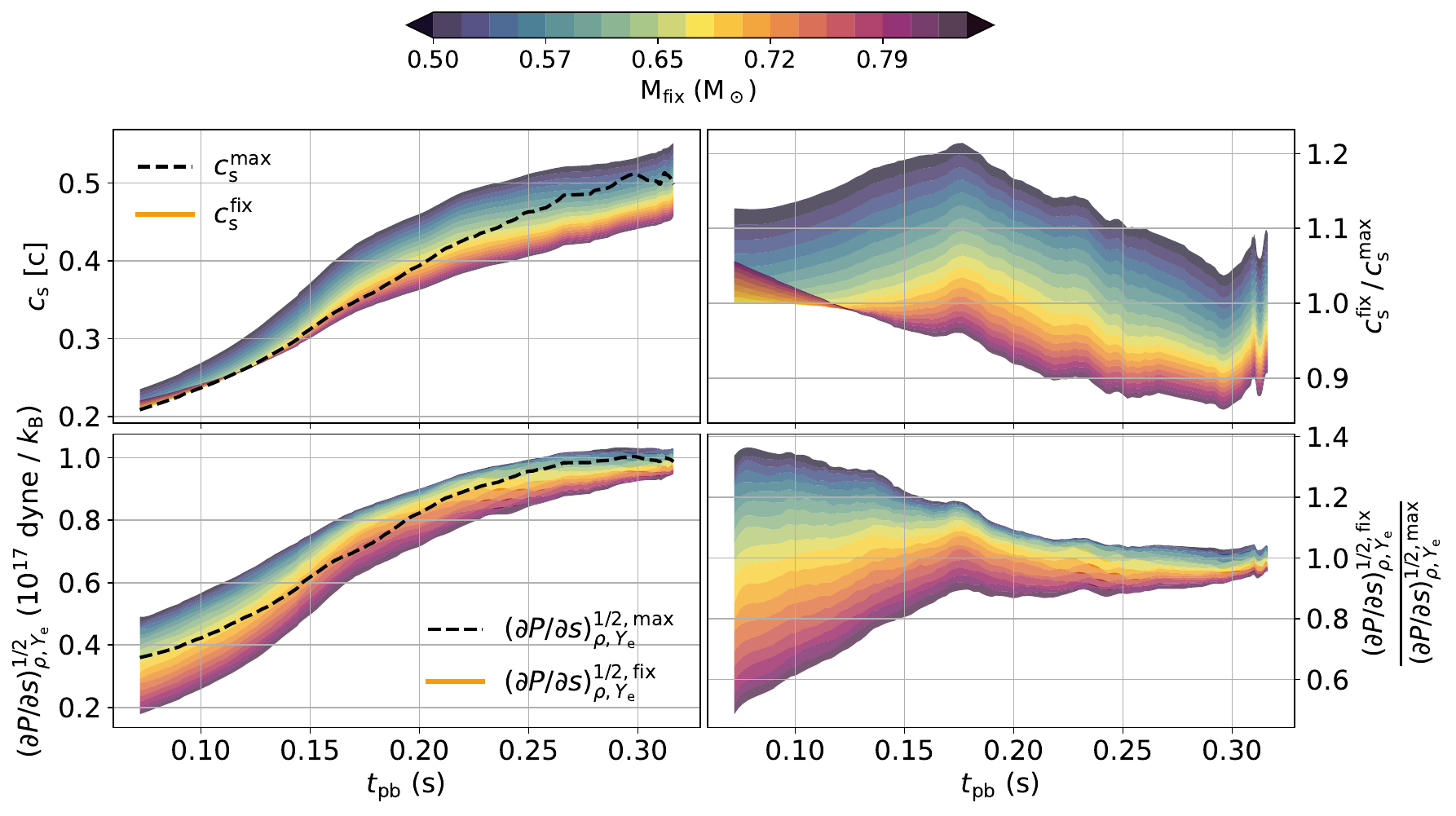}
\caption{Top row: Angle-averaged sound speed $c_\mathrm{s}$ at a fixed mass coordinate $M_\mathrm{fix}$ (coloured lines with colour bar at the top) for the 2D model \texttt{z85:CMF}. The black dashed line shows $c_\mathrm{s}$ at the peak of the Brunt-V\"ais\"al\"a frequency. Bottom row: Same plot for the square root of the thermodynamic derivative $\left(\partial P/\partial s\right)^{1/2}_{\rho,Y_\mathrm{e}}$. The  panels on the right show the corresponding {relative ratios} for different mass coordinates.}\label{fig:eos_terms}
\end{figure*}
We also notice that the maximum of the \bv is accompanied by a larger gradient $\D\rho/\D m$ (visible as more densely packed lines in the lower panel). Since the sound speed is highly sensitive to the density, $c_\mathrm{s}$ also exhibits a greater rate of change in this region ({see} the steepening of lines in the upper panel). The reason for the increased rate of change in density is the characteristic step in the entropy profile from shock heating shortly after bounce. 
%For a given pressure, density (usually) decreases with entropy, which is why the density exhibits a minor (reverse) step here too.  
The CMF models exhibit faster contraction, the trend for underestimating $c_\mathrm{s}^2$ is thus accentuated compared to \texttt{z85:SFHx}. This leads to an overestimation of the mode frequency at late times. The deviation is particularly evident in the spectrograms of Figure \ref{fig:spectra} (Panel 1) for \texttt{z35:CMF}, where the ``erroneously'' small sound speed (appearing in the denominator), and the fixed mass approximation (dashed black) together lead to a significant overestimation of the peak frequency. 
\begin{figure}
	\centering	\includegraphics[width=0.8\linewidth]{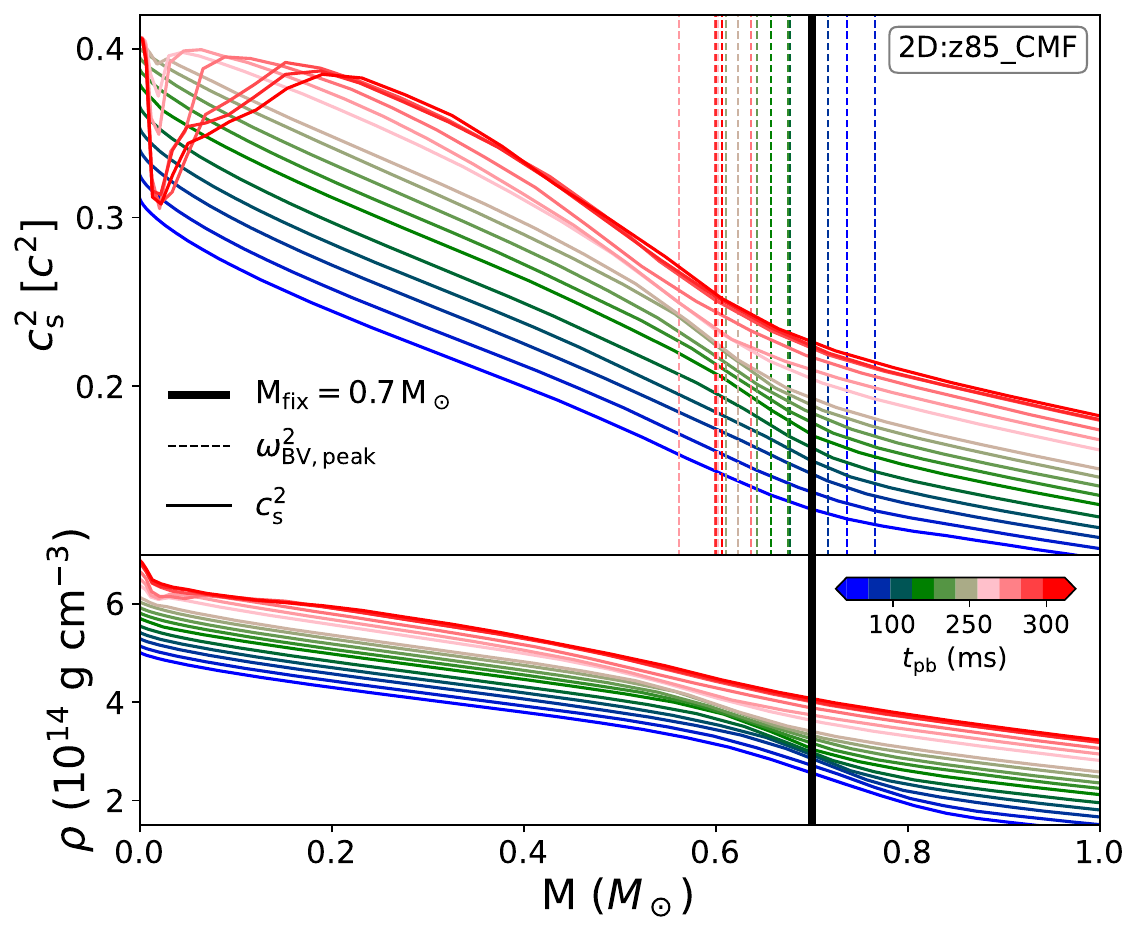}
\caption{Angle-averaged squared sound speed (top panel) as function of mass coordinate at different time steps (see bottom right colour bar)
for the 2D model \texttt{z85:CMF}. The coloured vertical lines indicate the maximum of $\omegabv$, tracked over time. The thick black line in the top panel shows the fixed mass coordinate $M=0.72\,\mathrm{M}_\odot$. Angle-averaged density profiles are shown in the bottom panel. We notice that the core $g$-mode region exhibits a steeper density profile (more densely packed lines, accompanied by a steeper radial speed of sound profile).}\label{fig:moving_gmode}
\end{figure}

\subsubsection{Approximation for the core $g$-mode frequency}
Putting together our approximations for $c_\mathrm{gr}$ and $c_\mathrm{eos}$, we obtain 
the analytic mode relation for the core $g$-mode,
\begin{align}\label{eq:bv_approx}
\tilde{\omega}_\mathrm{BV}^\mathrm{approx,fix}
&=
\nonumber
\frac{c_\mathrm{calib} \omega_\mathrm{BV}^\mathrm{approx,fix}}{2\pi}\\
&\approx 0.55\times \sqrt{\frac{1}{\pi}G M_\mathrm{mode}\alpha_\mathrm{approx}^5 
    \frac{1}{c_\mathrm{s}^2}\left(\pdv{P}{s}\right)_{\rho,Y_\mathrm{e}}\times 11.93\,\, k_\mathrm{B}/\mathrm{M}_\odot},
\end{align}
where we used an additional dimensionless calibration factor $c_\mathrm{calib}=0.55$.

We show our mode relation in the spectrograms of Figure~\ref{fig:spectra}. We present fits for the three 2D models of \citet{Jakobus_2023}, i.e., \texttt{z85:CMF}, \texttt{z35:CMF}, and \texttt{z85:SFHx} (from left to right).\footnote{No signal from the core $g$-mode was found for model \texttt{z35:SFHx}.}   
Each of the three columns shows,
\begin{enumerate}
\item GW spectrograms with overplotted lines for mode frequency fits.
    \item the effective relativistic acceleration appearing in the \bvns, $c_\mathrm{gr}~\equiv~\left[{\D\alpha}/{\D r}\!\cdot r^2 {\alpha}/{(h\phi^4)}\right]^{1/2}$,
    \item the EoS-factor $c_\mathrm{eos} \equiv \left(\partial{P}/{\partial s}\right)^{1/2}_{\rho,Y_e}/c_\mathrm{s}$,
    \item the radial entropy gradient $\D s/\D m$.
\end{enumerate}
\begin{figure*}
	\centering	\includegraphics[width=1\linewidth]{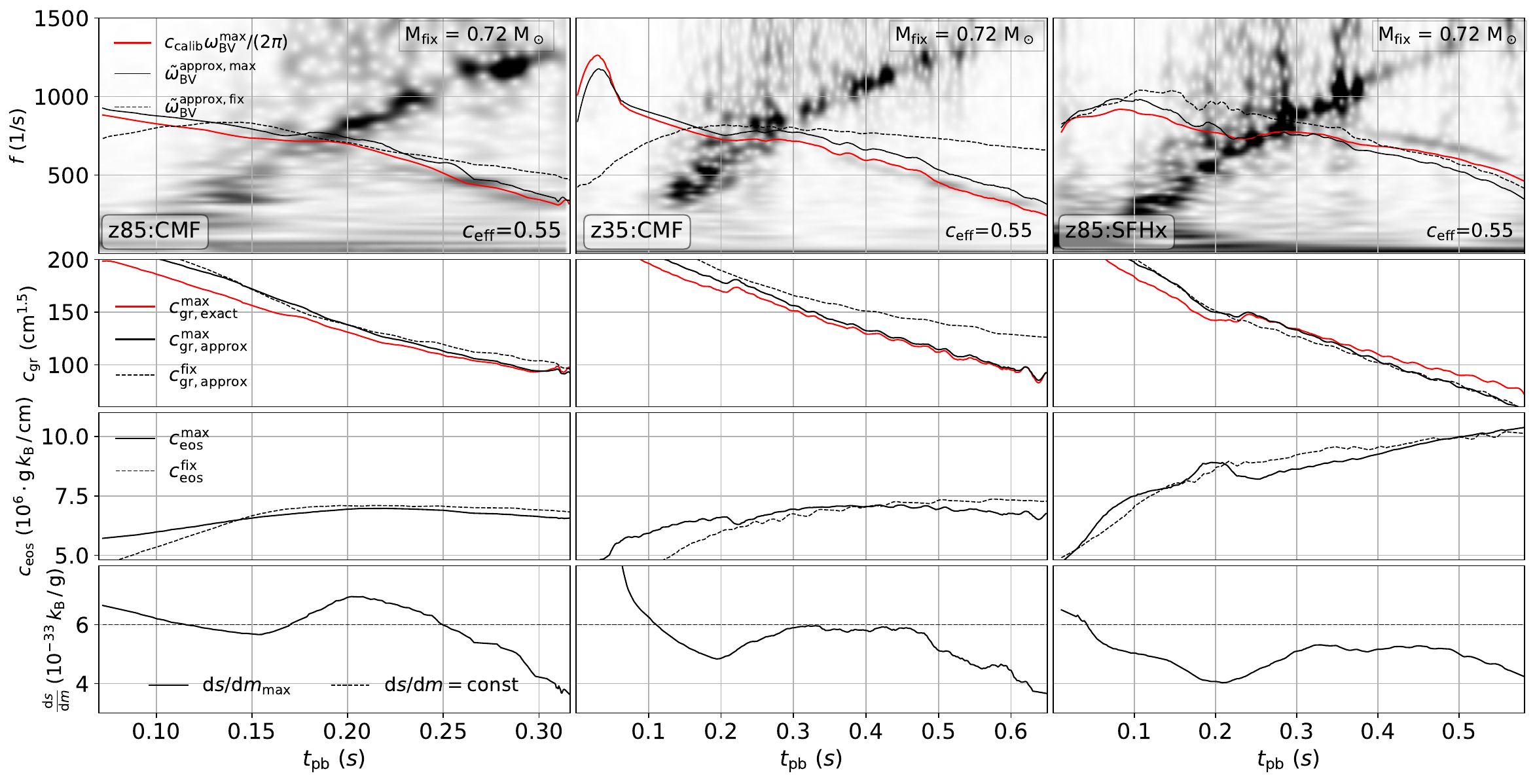}
\caption{First row: Spectrograms with overplotted mode fits for the 2D models \texttt{z85:CMF}, \texttt{z35:CMF}, and \texttt{z85:SFHx} (left to right). Red lines correspond to the square root of the actual Brunt-V\"ais\"al\"a frequency (divided by a factor $2\pi$). Black solid lines show the approximation $\tilde{\omega_\mathrm{BV}}^\mathrm{approx,max}$ from Equation~(\ref{eq:bv_approx}),
but evaluated at the actual peak of the \bvns. Black dashed lines 
show the frequency $\tilde{\omega}_\mathrm{BV}^\mathrm{approx,fix}$ evaluated at a fixed mass coordinate  $M_\mathrm{fix}$, which is displayed in the upper right corner of each panel. The second row shows the prefactor $c_\mathrm{gr}$, comparing the actual value of the
prefactor $c_\mathrm{gr,exact}^\mathrm{max}$ at the peak of the \bvns, the analytic approximation $c_\mathrm{gr,approx}^\mathrm{max}$ evaluated at the
actual maximum  of $\omegabv$, and the analytic approximation
$c_\mathrm{pre,approx}^\mathrm{fix}$ evaluated at
mass coordinate $M_\mathrm{fix}$. 
The third row shows the EoS factor $c_\mathrm{eos}$ at both the maximum of $\omega_\mathrm{BV}^2$ in solid black ($c_\mathrm{eos}^\mathrm{max}$), and at a fixed mass coordinate $M_\mathrm{fix}$
($c_\mathrm{eos}^\mathrm{fix}$). The last row compares the entropy gradient at the maximum of
$\omegabv$ (solid curve) to our approximation of a constant entropy gradient (dashed horizontal line).
}\label{fig:spectra}
\end{figure*}
The first row shows the actual peak value $f^\mathrm{peak}$ of the \bv, multiplied by a factor $c_\mathrm{calib}/(2\pi)$, in red lines for all three progenitors, with the simulation spectrograms shown in the background (in black and white). 
We further display the approximation from Equation~(\ref{eq:bv_approx}), evaluated at the actual peak of $\omegabv$ in black solid lines, and at fixed mass shell in black dashed lines, with the respective fixed mass coordinate indicated in the upper right corner of each spectrogram. We chose a fixed mass coordinate of $M_\mathrm{fix}=0.72\,\mathrm{M}_\odot$ for all three progenitors and $c_\mathrm{eff}=0.55$ (see Eq.~\ref{eq:ceff}). 
The approximation at $\omega_\mathrm{BW}^\mathrm{2,max}$ (black lines) and the exact form of $\omegabv$ (red lines) align fairly well (below $\mathord{\sim} 10\%$ relative error). We underestimate the mode frequency towards late times for progenitor \texttt{z85:SFHx}. The frequency approximation for \texttt{z35:CMF} is somewhat inaccurate, particularly at later times. Choosing a constant mass shell coordinate is the reason for this deviance as we will discuss below. 

The second row shows the relativistic prefactor $c_\mathrm{gr}$ (red) and our approximation at the maximum of the \bv (solid black). Analogously to the top row, the dashed black line shows the approximation at a fixed mass shell $M_\mathrm{fix}$ (dashed black).  The curves for the analytic approximations align fairly well in the $85 \mathrm{M}_\odot$ progenitor for both EoS (CMF and SFHx), but somewhat less so for the dashed curve in the top row that relies on the assumption of a fixed mass coordinate for the buoyancy jump.
By contrast, the lighter progenitor \texttt{z35:CMF} is clearly \textit{not} well captured by the fixed mass approximation (although the fit at the maximum \bv aligns very well). The reason for the misalignment is the direct dependence of the frequency on the square root of the core $g$-mode mass coordinate (see  $f^\mathrm{peak}\propto M_\mathrm{mode}^{0.5}$ in Eq.~\ref{eq:bv_approx}). Since the core $g$-mode mass coordinate shifts towards lower values as function of time, the mode is \textit{overestimated} at later times. {We find that our approximation also slightly overestimates the mode frequency in model \texttt{z85:CMF}
at late times, but the overestimation is less pronounced because the model does not run as long as \texttt{z35:CMF}.}

In the third row, we show the EoS parameter $c_\mathrm{eos}$. We compare values extracted at the peak frequency (solid black), and the ones obtained at mass coordinate $M_\mathrm{fix}$. Since we want to probe the EoS parameter, there is no approximation for the latter. 
The parameter $c_\mathrm{eos}$ is strongly sensitive to the location of the fixed mass shell. 

In the bottom row of Figure~\ref{fig:spectra}, we compare the time evolution of the entropy gradient
$\D s/\D m$ at the peak of the \bv to our approximation of a constant value for $\D s/\D m$. Clearly, $\D s/\D m$ is overestimated at later times. 
The decrease of $\D s/\D m_\mathrm{max}$ at late times is the result of neutrino diffusion and possibly convective overshoot into the stable right. Next to the specification of a fixed mass coordinate $M_\mathrm{fix}$, this marks the second largest error source in our analytic mode relation. 

\begin{table}
\centering 
\begin{tabular}{ll}
\toprule
$\alpha$ & $1 - G c^{-2}\left[{M_\mathrm{mode}}/{r_\mathrm{mode}} - {(M_\mathrm{NS} - M_\mathrm{mode})}/{\langle r\rangle}\right]$ \\
\midrule 
$\phi$ & $\alpha^{-1/2}$ \\
\midrule 
$h$ &  $\alpha^{-1/2}$\\
\midrule 
$\D\alpha/\D r$ & $G c^{-2} M_\mathrm{mode} h / (r_\mathrm{mode} \phi^2)^2$ \\
\midrule 
 $\D s/\D m$ & $11.93\,\, k_\mathrm{B}/\mathrm{M}_\odot$ \\
 \midrule 
$\left(\partial P/\partial\ye\right)_{\rho,s}\D\ye/\D r$ & neglect as $\ll \left({\partial P}/{s}\right)_{\rho,Y_\mathrm{e}}{\D s}/{\D r}$\\
\midrule 
$\left({\partial P}/{\partial s}\right)_{\rho,Y_\mathrm{e}}$ & EoS dependent parameter \\ 
\midrule 
$c_\mathrm{s}$ & EoS dependent parameter\\
\bottomrule 
\end{tabular}
\caption{Summary of approximations for the terms in the \bvns. The last two quantities, $\left(\partial P/\partial\ye\right)_{\rho,s}\D\ye/\D r$ and $c_\mathrm{s}$, are EoS parameters that need to be obtained for appropriate thermodynamic conditions based on a time-dependent solution of the PCS structure. }\label{table:approximations}
\end{table}

\section{Properties of the PCS convection zone}\label{sec:II}
In this section, we characterise the convective properties of the PCS. In particular, we focus on factors that may affect the excitation of the core ${}^{2}\!g_1$ mode and explain why the GW signal from the core $g$-mode is absent in our 3D simulation. For this reason, we restrict our analysis to the $85\,\mathrm{M}_\odot$ progenitor with the CMF EoS and compare results for the 2D and 3D simulations. 

\subsection{Explosion dynamics}\label{explosion_dynamics}
Before turning to PCS convection, we briefly discuss the dynamical evolution and explosion properties of the 2D and 3D models of the \texttt{z85:CMF} model. For this purpose, Figure~\ref{fig:expl_props} shows the the baryonic PCS mass $M$, the PCS radius $R$, the diagnostic explosion energy $E_\mathrm{expl}^\mathrm{diag}$, the mass accretion rate onto the PCS, and the angle-averaged shock radius $R_\mathrm{shock}$ for the 2D and 3D simulation.

Both the 2D and 3D models undergo shock revival. While the 2D model forms a black hole at a post-bounce time of $\mathord{\sim}330\,\mathrm{ms}$, the 3D model does not reach black hole formation before the end of the simulation at $\mathord{\sim}430\,\mathrm{ms}$. In both models, the explosion dynamics are comparable to the \texttt{z85} run in \cite{powell_21}, which was, however, performed with the \texttt{SFHx} EoS. In the 3D simulation, the baryonic PCS mass is slightly increased, reaching $\mathord{\sim}2.55\,\mathrm{M}_\odot$, along with a decreased mass accretion rate. The PCS radius is predominantly determined by the PCS EoS and reaches similar final values of $\mathrm{\sim}24-25\,\mathrm{km}$, but PCS contraction is slightly faster in 2D. Note that there are some early differences in the PCS mass, mass accretion rate, and shock trajectory right after bounce. We have traced these to slightly different collapse
dynamics in both models. To avoid artefacts from slightly imperfect matching between the low-density EoS and the SFHx EoS at intermediate densities, the switch between those EoS regimes was set at a higher density in the 2D model during the initial collapse phase than in the 3D model. The transition density was then set to the same value before significant electron fraction and entropy changes occur later in the collapse, but this imprinted small entropy differences between 2D and 3D in parts of the core, and resulted in slightly different collapse times of
$0.43\,\mathrm{s}$ in 3D and $0.49\,\mathrm{s}$ in  2D. The mass of the homologous core differs slightly between 2D ($0.7\,\mathrm{M}_\odot$) and 3D ($0.67\,\mathrm{M}_\odot$) as do the post-bounce entropy profiles, and this then also affects the dynamics of prompt convection.
These differences, compounded by the stochasticity, preclude a perfect comparison of the 2D and 3D model and illustrate once again a non-negligible sensitivity of the post-collapse phase to the collapse physics \citep[cp.][]{lentz_12}.

In the 2D simulation, shock revival occurs at $\mathord{\sim}150\,\mathrm{ms}$, whereas in the 3D simulation, it occurs slightly later at $\mathord{\sim}180\,\mathrm{ms}$. The diagnostic explosion energy, which we compute in general relativity following \citep{mueller_12a}, reaches comparable values in both cases. The explosion energy in these two CMF models is higher than that of the 3D model with the SFHx EoS computed by \citet{powell_21}.
\begin{figure}
	\centering	\includegraphics[width=0.8\linewidth]{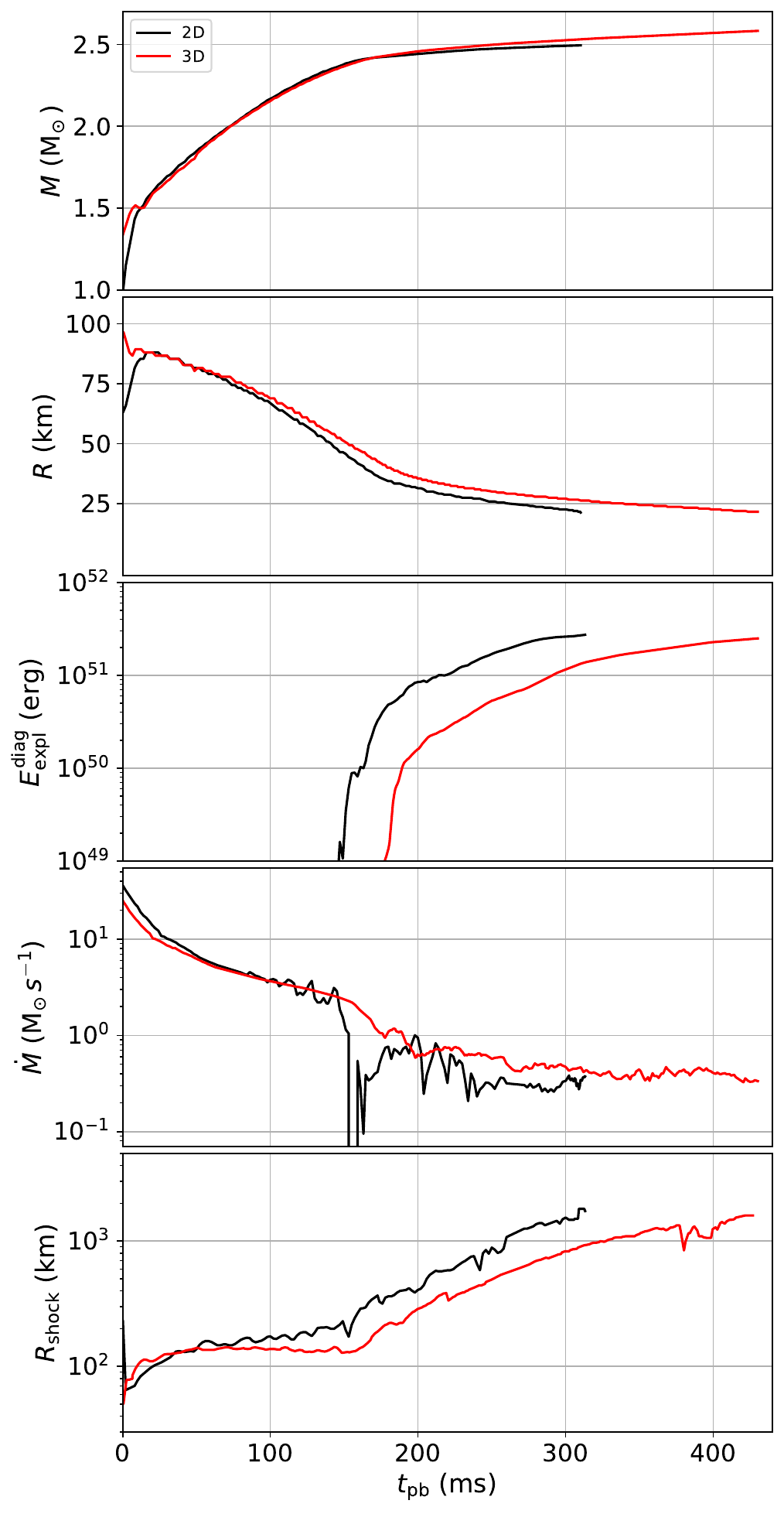}
\caption{Time evolution of PCS mass (first row), PCS radius (second row), diagnostic explosion energy (third row), mass accretion rate (fourth row), and angle-averaged shock radius (fifth row). Both models undergo shock revival within $155-180\,\mathrm{ms}$ after core bounce. The 3D model does not reach black hole formation before the end of the simulation.}\label{fig:expl_props}
\end{figure}
\subsection{Time variability of the quadrupole moment}\label{subsec:time_variab}
  
We observe a strong core $g$-mode signal for model \texttt{2D:z85}, but not for model \texttt{3D:z85}.  
There are three possible reasons for the less efficient excitation of the core $g$-mode by PCS convection in 3D.
The turbulent kinetic energy in the convection zone may be smaller, less of the turbulent kinetic energy may be contained in quadrupole motions, and the frequency spectrum of convective motions may overlap less with the mode frequency.
We first investigate the possibility of a smaller turbulent kinetic energy in the PCS convection zone in 3D. 
To this end, we compute the turbulent kinetic energy inside the PCS, defined as 
\begin{equation}\label{eq:ekin_turb}
       E_\mathrm{kin,turb}^\mathrm{PCS} = \frac{1}{2}\int_\mathrm{\rho\geq 10^{11}\,\mathrm{g\,cm}^{-3}}\rho v_\mathrm{turb}^2 \mathrm{d}V, 
\end{equation}
where $v_\mathrm{turb} =  v_r^{'2} + v_\theta^2 + v_\phi^2$ and $\mathrm{d}V = \phi^6 r^2 \sin\theta  \,\mathrm{d}\phi\,\mathrm{d}\theta$ in general relativity. The quantity $v_r^{'}$ is the Favrian radial velocity perturbation, which is obtained from the Favre (i.e., density weighted) average   $\langle v_r\rangle$ as $v_r^{'} = v_r - \langle v_r\rangle$
. The turbulent kinetic energy as a function of post-bounce time for the 3D and 2D simulations is shown in Figure~\ref{fig:ekinturb}. In both models, the turbulent kinetic energy saturates at similar values of approximately $8\times 10^{50}\,\mathrm{erg}$. In the 3D simulation, however, the increase occurs abruptly, around $180\,\mathrm{ms}$ after bounce, compared to 2D, where the turbulent kinetic energy builds up more slowly within $\mathord{\sim} 100\,\mathrm{ms}$. Hence the absence of the core $g$-mode signal in 3D cannot be ascribed to weaker convective forcing.

We therefore investigate the second potential explanation, i.e., that a lack of power in $l=2$ perturbations could explain the less efficient excitation of the core $g$-mode excitation.
\begin{figure}
	\centering	\includegraphics[width=0.7\linewidth]{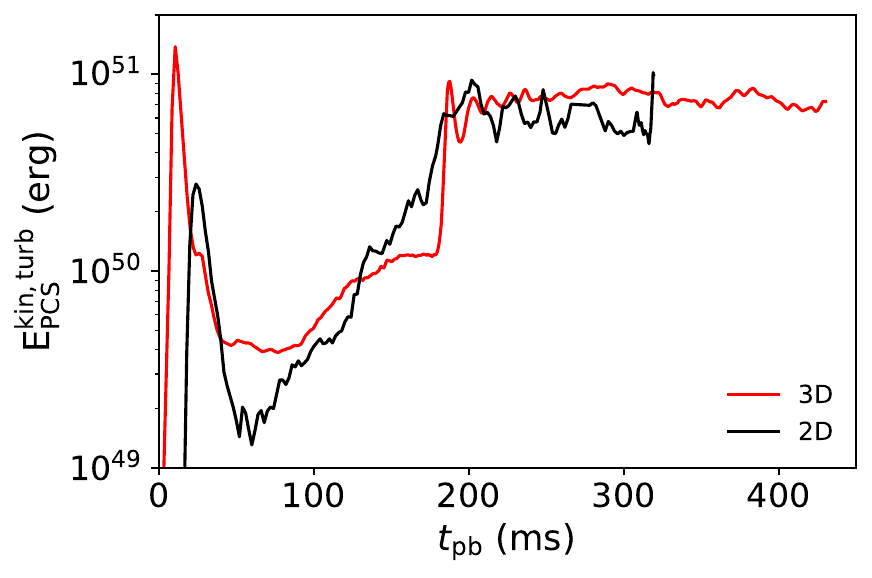}
\caption{Time evolution of the PCS turbulent kinetic energy for 2D (black line) and 3D (red line); see Equation~(\ref{eq:ekin_turb}) for the definition of the turbulent kinetic energy. Both models exhibit similar turbulence strengths in the PCS. The 3D run has a sudden ``step'' at $\mathord{\sim} 180\,\mathrm{ms}$, during which the turbulent kinetic energy rises sharply.}\label{fig:ekinturb}
\end{figure}
For this purpose, we compare the turbulent energy spectra $E(l)$  as function of the multipole order $l$ in 2D and 3D at two different post-bounce times in the first row of Figure~\ref{fig:decomp_turb}. We decompose the radial turbulent convective motion into spherical harmonics $\mathcal{Y}_\mathrm{lm}(\phi,\theta)$ according to
\begin{align}
     E(l) = \frac{1}{2}\sum_{m=-l}^l \bigg\rvert \int \mathcal{Y}_{lm}^* v_r \sqrt{\rho} \,\mathrm{d}\Omega\bigg\rvert ^2, \label{eq:spectral}
\end{align}
where we sum over modes corresponding to the same multipole order $l$. To obtain smoother spectra, we average over five radial zones in the 3D run and 30 radial zones in the 2D run, and average over a time window of approximately $10\,\mathrm{ms}$. 

In Kolmogorov's theory of turbulence, the energy spectrum is determined by a forward energy cascade to small scale with a $E(l) \propto l^{-5/3}$ scaling law in the energy spectrum~\citep{Kolmogorov_1941}. In this case, turbulent energy is injected by external sources at the largest (spatial) scales and cascades to smaller scales. Conversely, turbulence in 2D undergoes an ``inverse turbulent energy cascade'' wherein turbulent energy is transferred from small scales to large scales~\citep {Kraichnan_1967}. These different behaviours significantly affect the post-shock dynamics~\citep{Murphy_2011, Hanke_2013}.

In Figure~\ref{fig:decomp_turb}, we show the $5/3$ power-law as a green line, and the $l^{-3}$ power-law as dashed green. 
The power spectrum observed in the 3D run aligns well with the predictions of turbulent theory for length scales $l\gtrsim (15\texttt{-}40)$ where energy is transferred to small scales until dissipation becomes prominent at $l\approx 40$. 
In 2D, the spectrum fluctuates significantly, making it difficult to identify a clear power-law trend. At smaller scales $l\gtrsim 15$ the 2D run roughly follows a $l^{-3}$ forward enstrophy cascade, see dashed green line \citep{Kolmogorov_1941,Kraichnan_1967,Hanke_2012}. Based on these snapshots, it is difficult to judge, however, whether quadrupolar motions are generally stronger in 2D.

Better insights can be gained from the second panel of Figure~\ref{fig:decomp_turb}, which presents the time evolution of the power $E(2)$ in quadrupole motion in the 3D run (dotted red lines) and the 2D run (dotted black lines). The turbulent quadrupole motion exhibits comparable magnitudes in the 2D and 3D simulations on average, but we observe a significantly higher level of temporal variability in the 2D run.
Additionally, in the 2D run, we observe enhanced power at lower scales with $l\lesssim 10$ and a relatively smaller proportion of power at small scales, compared to the total power $E_\mathrm{tot}$ (solid lines, see distance of solid lines versus dashed lines). This is a consequence of the $l^{-3}$ scaling law in 2D.  
\begin{figure}
	\centering	\includegraphics[width=1\linewidth]{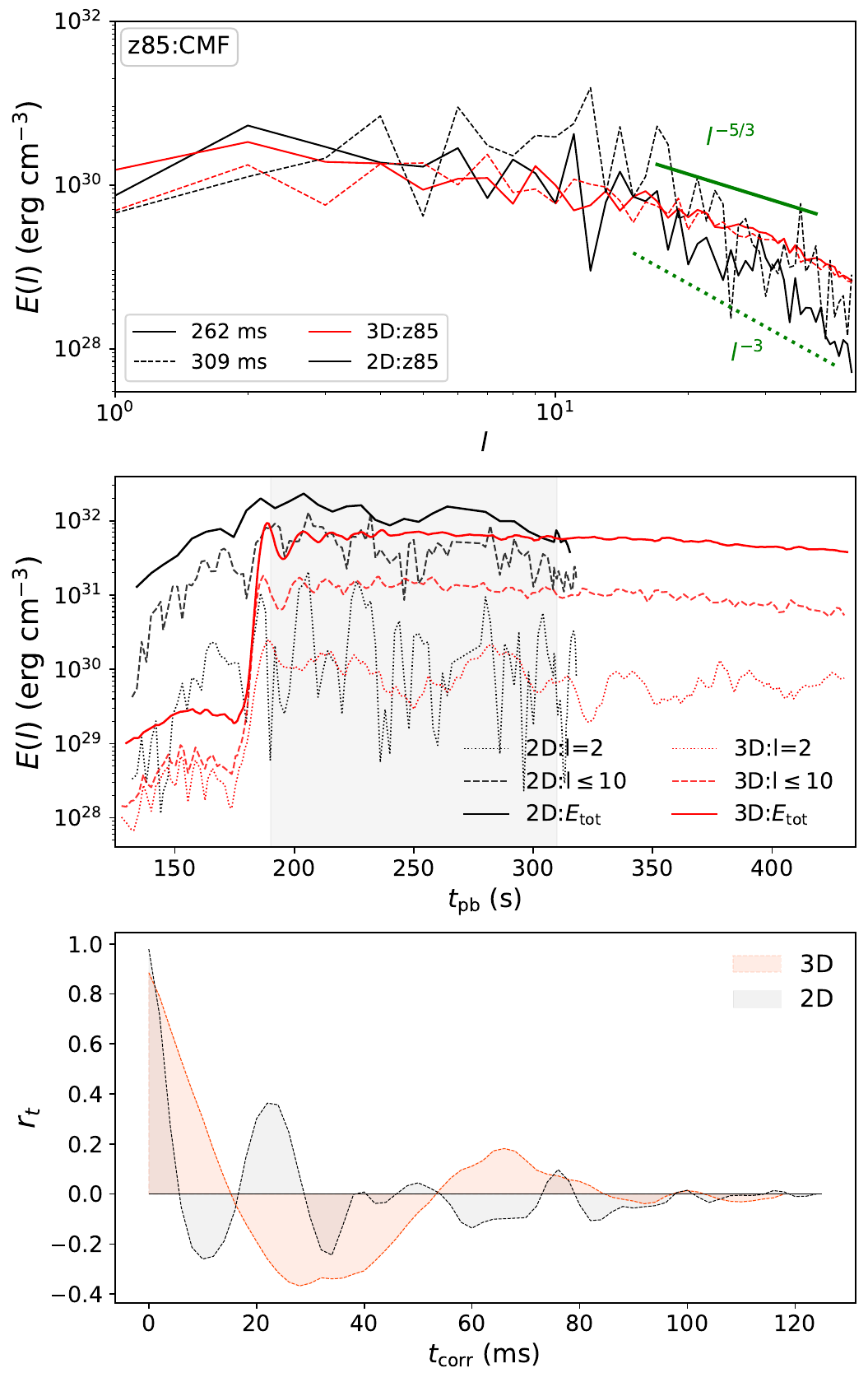}
\caption{Top panel: Turbulent energy spectrum $E(l)$ from Equation~(\ref{eq:spectral}), evaluated at $13\,\mathrm{km}$ for the 3D run (red), and at $11\,\mathrm{km}$ for the 2D run (black) at two different times as indicated in the legend in the upper right. We choose a slightly smaller radii in 2D, as the convective zone is located further inside compared to 3D. Middle panel: Time evolution of different components of the turbulent energy spectrum (3D: red,  2D: black), i.e., the quadrupolar power $E(2)$, the power at $l\leq 10$, and the total power
$E_\mathrm{tot}$ over the entire spectrum evaluated as well at 13 km for the 3D run, and 11 km for the 2D run. Bottom panel: Autocorrelation function of the quadrupole spectral power $E(2)$ in 3D (red shaded) and 2D (black shaded) as computed from Equation~(\ref{eq:acf}). The time interval for the computation of the autocorrelation function is indicated as the grey-shaded area in the middle panel. The first panel shows increased variability in the power spectrum as a function of different multipole order $l$ in 2D compared to 3D. We further observe increased time variability of the quadrupolar moment in 2D (middle panel) and a shorter autocorrelation time (bottom panel).}\label{fig:decomp_turb}
\end{figure}
It should be noted here that, although $E_\mathrm{tot}$ is marginally higher in the 2D run, the turbulent energies $E_\mathrm{kin,turb}^\mathrm{PCS}$ in Figure~\ref{fig:ekinturb} are of similar order.\footnote{We evaluate the total energy at a given radius via the non-decomposed form of Equation~(\ref{eq:spectral}), e.g. $E_\mathrm{tot} = \frac{1}{2}\int\rho v_r^2 \mathrm{d}\Omega$. Thus, this quantity is not to be confused with the total turbulent kinetic energy in the PCS convection zone.} 

To further quantify the time variability of PCS convection in 2D and 3D, we plot the autocorrelation function of
$E(2)(t)$  in the third panel of Figure~\ref{fig:decomp_turb}.    
The autocorrelation function $r(t)$ of a function $E(t)$ is given by 
\begin{equation}\label{eq:acf}
r(t) = \int E(t+\tau) E(\tau)\, \mathrm{d}\tau,  
\end{equation}
which we evaluate using a discrete time series with sampling rate $\tau = 10^{-3}\,s$ for $E(2)(t)$.
To avoid non-stationary trends, we use a time interval in which the quadrupolar strength is roughly constant in between $\mathord{\sim}180 \texttt{-} 310\,\mathrm{ms}$ for both 2D and 3D. The grey shaded areas in the second panel of Figure~\ref{fig:decomp_turb} indicate the time interval.
It is evident that the autocorrelation time $\tau_\mathrm{corr}$ (which we define as the first zero of the correlation function) in the 2D run is notably shorter compared to the 3D run.
This disparity between 3D and 2D could be attributed to factors such as faster decay rates of eddies or a greater velocity dispersion around the typical convective velocity within the turbulent flow, or simply by greater temporal variability in a 2D system in which a smaller number of modes has to carry roughly the same convective flux as in 3D.

To further assess the stronger temporal variability of quadrupolar motions in 2D as a factor for the emergence of the core $g$-mode signal, one could also consider (temporal) Fourier transforms of the turbulent energy spectrum.
The Wiener-Khinchin theorem relates the Fourier transform and the autocorrelation function~\citep{Khintchine_1934}
\begin{equation}\label{eq:khinchin}
    r(t) = \mathcal{F}_f \left(|(\tilde{E}(f )|^2)\right)(t),
\end{equation}
where $\tilde{E}(f) = 1/\sqrt{2\pi}\int E(t) e^{-2\pi i f t}\D t$ is the Fourier transform of $E(t)$ and $\mathcal{F}_f$ denotes the Fourier transform of the power spectral density $|(\tilde{E}(f )|^2$. Contrary to the Fourier transform, the autocorrelation has no phase information (due to the magnitude-squared operation), so it is not possible to invert Equation~(\ref{eq:khinchin}) to obtain the Fourier transform $\tilde{E}(f)$. The retention of phase information in the Fourier transform comes at the price of a rather noisy signal in the frequency domain compared to the rather smooth autocorrelation function, however.

Nonetheless, it is instructive to compute the temporal Fourier transforms of the turbulent energy spectrum $E(l)$ for $l\leq 10$ in the post-bounce time window $180 \,\mathrm{ms} \mathord{\leq} t \mathord{\leq} 310\mathrm{ms}$ (Figure~\ref{fig:fourrier})
in 3D (upper panel), and 2D (lower panel). We note the increased power at higher frequencies in 2D, which once more reflects the stronger variability of convection on short time scales in 2D, as already pointed out in our analysis of the correlation time for the quadrupole moment. 
\begin{figure}
	\centering	\includegraphics[width=0.8\linewidth]{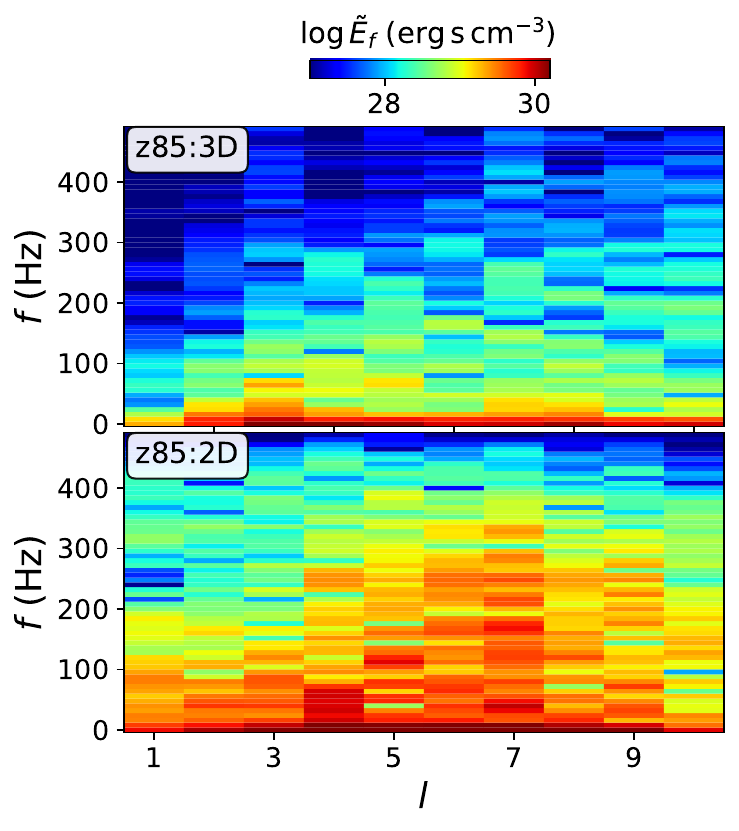}
\caption{Temporal Fourier transform of the turbulent energy spectrum $E(l)$ for $l\leq 10$, computed for the same time interval as the autocorrelation function in Figure~\ref{fig:decomp_turb}, i.e., during the window $\mathord{\sim}180 \texttt{-} 310\,\mathrm{ms}$. More power at high frequency is observed in the 2D run, reflecting stronger temporal variability of the convective energy in quadrupolar motions on short time scales.}\label{fig:fourrier}
\end{figure}

To wrap this subsection up, we briefly summarize our findings: In Fig.~\ref{fig:decomp_turb}, the decomposed turbulent motion in the PCS convection zone exhibits a  $l^{-5/3}$ scaling law at intermediate harmonic degrees $l\gtrsim 15$ in 3D, and a steeper $l^{-3}$ decline in 2D at intermediate degrees $l\gtrsim 15$, which is both in accordance with turbulent theory~\citep{Kraichnan_1967, Kolmogorov_1941,Landau_1959}. The inverse cascade in 2D is less established at small $l\lesssim 5$, and 2D and 3D show similar behaviours here. There is more power though at low, and intermediate harmonic degrees $5\lesssim l\lesssim 15$ in 2D, as we would expect from an inverse cascade (this is more so the case at later postbounce time $\mathord{\sim} 300\,\mathrm{ms}$). The second thing we notice is a large temporal variability of the turbulent quadrupole in 2D compared to 3D (second panel of Fig.~\ref{fig:decomp_turb}). This is reflected in a shorter correlation time of the time evolution of $E(2)$ (third panel), and as increased power at higher frequencies in the Fourier spectra $\tilde{E}(l)$ in Fig.~\ref{fig:fourrier}.

We can thus summarise our analysis of the possible reasons for weaker excitation of the core $g$-mode as follows.
As shown in Figures~\ref{fig:ekinturb} and \ref{fig:decomp_turb}, the total turbulent kinetic energy and the energy in quadrupolar motions are not significantly reduced in 3D compared to 2D in a time-averaged sense. The most likely explanation for the lack of the core $g$-mode signal in 3D is therefore the increased correlation time, smaller temporal variability, and smaller power at high frequencies in convective motions in 3D.
This analysis supports earlier qualitative reasoning \citep{andresen_2017} that the eddies in the 2D simulation exert more ``impulsive'' forcing and that their frequency spectrum overlaps to a higher degree with the natural core $g$-mode frequency, allowing for resonant excitation of the core $g$-mode oscillation. 
{It is important to note that the reduced high-frequency
forcing from PNS convection will affect the core and surface $g$-mode signals differently. Since the surface $g$-mode lives at high frequency, one might superficially expect it to be damped more strongly in 3D, but even if this is the case, the core $g$-mode is more likely to fall below the noise threshold simply because it is already weaker in 2D to begin with. Moreover, different from the core $g$-mode, the surface $g$-mode is
not just excited by PCS convection, but also by downflows from above each contribute to exciting the surface g-mode \citep{Murphy_2009,Marek_2009,mueller_2013,yakunin_2015}.
Furthermore, the excitation of the modes depends not only on the
frequency spectrum of the forcing, but also on the spatial structure of the eigenmodes and the structure of the convective boundary
and on the impinging convective energy flux available to drive
$g$-mode excitation (e.g., \cite{Lighthill_1967,goldreich_90,Aerts2016}). The convective energy flux
is very different at the upper and lower boundary of the PCS
convection zone (see Section~\ref{sec:conv_velocities}). For these
reasons, the relative power in the core and surface $g$-mode is
ultimately \emph{not} decided by which mode matches the frequencies of
PCS convection better -- the surface $g$-mode is always stronger empirically. But the different frequency spectrum between 2D and
3D at a \emph{given} convective boundary will still influence whether
mode excitation is stronger or weaker in 3D.}

\subsection{Energy fluxes in the PCS convection zone}\label{sec:conv_velocities}
As discussed in the previous section, the turbulent kinetic energy is a determining factor for the excitation of oscillation modes by PCS convection and the GW emission by these modes. For convection in the gain region, one can directly relate the convective energy flux to the power in GWs
\citep{powell_2019,Radice_2019}.
A similar relation has been postulated
between the neutrino luminosity and
GWs from modes excited by PCS convection \citep{mueller_17gw} by invoking a balance between the neutrino luminosity and the PCS convective energy flux. It is unclear, however, how well such balance arguments hold for PCS convection.
In other contexts, e.g., for convection
during the neutrino-cooled burning stages of massive stars, such balance arguments have been discussed extensively \citep[e.g.,][]{mueller_2016}.
For convection driven by thin burning shells, the dominant source term must balance convective energy transport and turbulent dissipation so that the net entropy generation rate in the convective region is roughly uniform and a build-up of a growing unstable entropy gradient is avoided.  
In the context of PCS convection, one might expect a similar self-regulation of the convective flux.
When the convective region in the PCS cools due to the total neutrino energy flux $F_\mathrm{rad}^\nu$, the entropy gradient will be pushed towards negative values; the magnitude of the negative gradient will depend on the neutrino flux $F_\mathrm{rad}^\nu$ at the outer boundary of the convective zone.
A steeper negative entropy gradient will enhance entropy-driven convection, causing hotter material from the lower layers to be transported towards the outer boundary of the convective region, resulting in a large convective energy flux $F_\mathrm{conv}^\mathrm{h}$. This counteracts the build-up of a negative entropy gradient until balance is achieved. 
\begin{figure*}
	\centering	\includegraphics[width=0.8\linewidth]{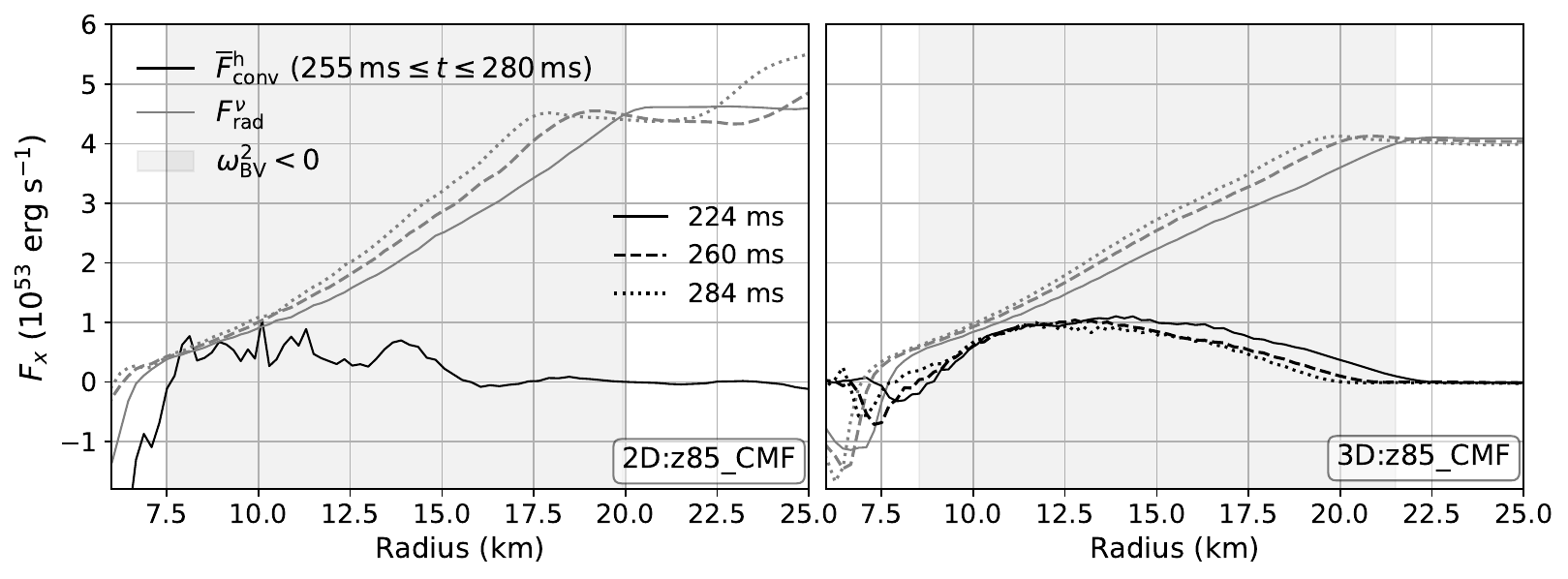}
\caption{Total neutrino energy flux $F^\nu_\mathrm{rad}$ (light grey lines) at three different times with corresponding legend in the left plot, and convective energy flux $F_\mathrm{conv}^\mathrm{h}$ (black) in 2D (left) and 3D (right). The convective energy flux in 2D is averaged over $25\,\mathrm{ms}$ between $255\,\mathrm{ms} \texttt{-} 280\,\mathrm{ms}$ after bounce. In both models, the total neutrino energy flux $F^\nu_\mathrm{rad}$ exceeds the convective flux $F_\mathrm{conv}^\mathrm{h}$ towards the outer convective boundary.}\label{fig:flux3}
\end{figure*}
But although some form of balance must still govern the quasi-steady state of PCS convection, the driving of convection involves a more complex interplay of competing lepton and entropy gradients, and the relation between the bulk energy loss (i.e., the total neutrino luminosity) and the strength of convection may be complicated.
In this section we will analyse what factors determine the turbulent convective energies in the PCS convection zone, e.g., whether the convective velocities can be related to the total outgoing neutrino flux.

In our analysis we shall make repeated use of spherical Reynolds and Favre averaging.
We denote volume-weighted spherical Reynolds averages as $\langle.\rangle$ or hats for single averages, and mass weighted Favre averages as tildes $\tilde{X} = {\langle\rho X \rangle}/{\hat{\rho}}$ \citep{favre_1965}. We use single primes for fluctuating quantities from Reynold averages $X' = X - \hat{X}$, and double primes for fluctuations from Favre averages ${X}'' = X - \tilde{X}$.

To analyse the quasi-steady state of PCS convection, we plot the total neutrino energy flux $F_\mathrm{rad}^\nu$, and convective energy flux $F_\mathrm{conv}^\mathrm{h}$ both 2D and 3D in Figure~\ref{fig:flux3}. 
We  define the total neutrino energy flux $F_\mathrm{rad}^\nu$ as
\begin{align}
    F_\mathrm{rad}^\nu & = 4\pi r^2 \langle F_{\nu_\mathrm{e}} + F_{\overline{\nu}_\mathrm{e}} + 4F_{\nu_\mathrm{x}}\rangle \nonumber \\
     &= \int W\alpha\phi^4 (F_{\nu_\mathrm{e}} + F_{\overline{\nu}_\mathrm{e}} + 4F_{\nu_\mathrm{x}}) r^2\D\Omega, \label{eq:flux_rad}
\end{align}
where $W$ is the Lorentz factor, $\alpha$ the lapse function, and $\phi$ the conformal factor, and $F_{\nu_\mathrm{e}}$,
$F_{\bar{\nu}_\mathrm{e}}$, and $F_{\bar{\nu}_\mathrm{x}}$
are the energy flux densities of electron neutrinos, electron antineutrinos, and heavy-flavour neutrinos, respectively. The convective energy flux is defined as  
\begin{align}
    {F}_\mathrm{conv}^\mathrm{h} &= 4\pi r^2  \langle\rho {v}_r'' {h}'' \rangle= \int \alpha\phi^4 \rho{v}_r''{h}''r^2\mathrm{d}\Omega. \label{eq:conv_h}
\end{align}
Both fluxes are plotted at three different time steps, as indicated in the centre right of the left panel. The 2D convective energy flux is averaged over those three time steps since it fluctuates strongly.

Notably, we do not observe
a balance ${F}_\mathrm{conv}^\mathrm{h}\approx F_\mathrm{rad}^\nu$
in Figure~\ref{fig:flux3}. The neutrino energy flux is, in fact, multiples times larger than the convective energy flux, particularly towards the outer PCS convection zone, i.e., from the viewpoint of energy transport PCS convection is inefficient, somewhat akin to the outer layers of stellar surface convection zones \citep{kippenhahn_1994}. This means that a crude estimate 
of the convective velocity by dimensional analysis as $v_\mathrm{conv}\sim 
[{F}_\mathrm{conv}^\mathrm{h}/(4\pi r^2 \rho)]^{1/3}
\sim[{F}_\mathrm{rad}^\mathrm{\nu}/(4\pi r^2 \rho)]^{1/3}
$ \citep{mueller_2016,mueller_17gw} will be systematically too high.
The key difference compared, e.g., to the late burning stages in massive shells that explains this behaviour is the more important role of radiative diffusion throughout the convective shell and the occurrence of cooling over a more substantial fraction of the convection zone rather than just a thin layer at the boundary. This implies that one needs to consider a more complex balance for the local entropy source and sink terms
$\dot{s}_\mathrm{conv}$ and $\dot{s}_\nu$
from convective and radiative energy transport, and from convective and diffusive net lepton number transport, instead of simply equating the maximum of $\mathbf{F}_\mathrm{turb}$ to the neutrino flux from the PCS convection zone. 
{To obtain the entropy source and sink terms we therefore also need to introduce the } convective lepton number flux\footnote{As mentioned by~\citet{powell_2019}, the convective lepton number flux in our simulations does not include any advective neutrino flux since the FMT scheme does not include co-advection of neutrinos with matter.},
\begin{align}\label{eq:leptonflux}
{F}_\mathrm{conv}^\mathrm{lep} &=  4\pi r^2 \langle\rho {v}'' {Y}''_\mathrm{e} \rangle
\nonumber
\\
&= \int W\alpha\phi^4 \rho{v}_r''{\ye}''r^2\D\Omega,
\end{align}
We further compute the diffusive net lepton number flux
\begin{align}
{F}^\mathrm{lep}_{\nu} &=  \int W\alpha\phi^4 (\mathcal{F}_{\nu_e}- \mathcal{F}_{\overline{\nu}_e})r^2\D\Omega,\label{eq:diffusive_net_lepton}
\end{align}
where $\mathcal{F}_{\nu_i} = F_{\nu_i}/\langle E \rangle_{\nu_i}$ is the electron neutrino and anti-neutrino number flux density, computed from $F_{\nu_i}$ and the average neutrino energy
$\langle E \rangle_{\nu_i}$. 

One expects that the local rate of change 
\begin{align}
\dot{s}_\mathrm{conv}+\dot{s}_\nu & \approx 
\label{eq:balance1}
-\frac{m_\mathrm{B}}{4\pi k_\mathrm{B}\rho T r^2 \phi^6}
\frac{\partial}{\partial r} \left[\alpha \phi^4\left(
 F_\mathrm{conv}^\mathrm{h}+
F_\mathrm{rad}^\nu\right)\right] 
\\ 
&- \frac{\mu_\nu m_\mathrm{B}}{4\pi k_\mathrm{B} \rho T r^2 \phi^6}
\frac{\partial}{\partial r} \left[\alpha \phi^4\left(
F_\mathrm{conv}^\mathrm{lep}+
F_\nu^\mathrm{lep}\right)\right],
\nonumber
\end{align}
should be approximately constant to avoid a secular build-up of an entropy gradient.
\begin{figure}
	\centering	\includegraphics[width=\linewidth]{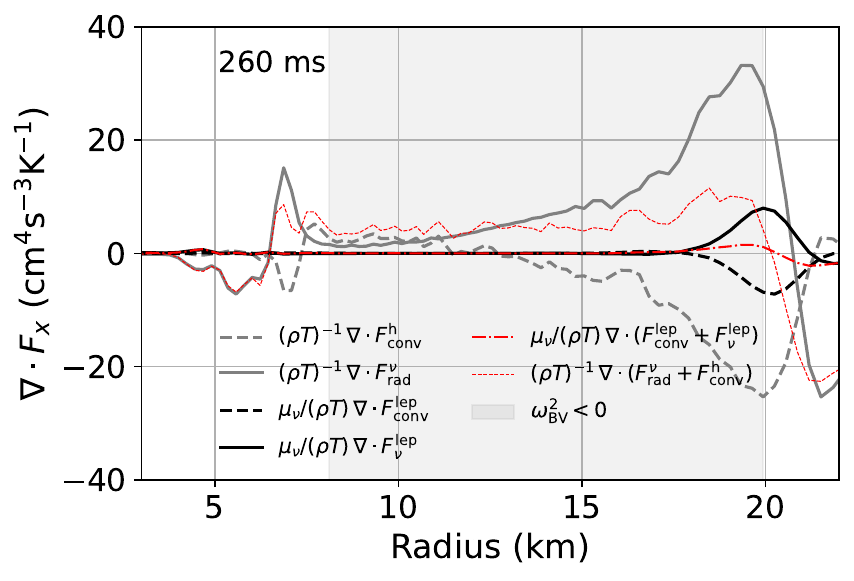}
\caption{Divergences of convective energy flux (grey), total neutrino energy flux (dashed grey), convective lepton flux times neutrino chemical potential (black dashed), and diffusive net lepton number flux with neutrino chemical potential (black), all as function of radius at 260 ms for the 3D model. 
For brevity, we omit relativistic prefactors in the legend of this plot, which however enter the expressions for the divergences, see Equation~(\ref{eq:balance1}). }\label{fig:flux_div}
\end{figure}
We evaluate these local entropy source and sink terms in
in 3D\footnote{The profiles of the convective flux are too noisy for this analysis in 2D.} in Figure~\ref{fig:flux_div}, where we plot the divergences of the convective energy flux (dashed grey), total neutrino energy flux (grey line), convective lepton number flux (dashed black), and the diffusive net lepton number flux (black line). Their respective (partial) sums are shown as a thick dashed red line (sum of the divergences of lepton number fluxes) and as a dash-dotted red line (sum of the divergence of energy fluxes). 
The divergence of the convective energy flux somewhat counteracts the divergence of the total neutrino energy flux; the sum of both terms is shown as a dotted red line which is approximately constant albeit slightly increasing towards the outer convective boundary. 
The divergences of convective lepton number flux and diffusive net lepton number flux almost cancel each other, although their sum becomes slightly positive towards the outer convective boundary. 
Since the sum of these two is approximately zero within the PCS convection zone (dash-dotted), it is apparent by eye that the sum of all four flux divergence terms is approximately constant, with the divergences of the energy fluxes in the two terms in Equation~(\ref{eq:balance1}) being the dominant terms.
Thus, energy transport in the PCS convection zone is well described by a secular balance condition.
Unfortunately, this condition of local balance
$\dot{s}_\mathrm{conv}+\dot{s}_\nu \approx \mathrm{const.}$ does not lend itself easily to a simple estimate for the typical convective velocity any more.

It is still worthwhile to consider, however, whether one could obtain reasonable convective velocities if
$F_\mathrm{conv}^\mathrm{h}$ were known (e.g., from mixing-length theory or from some non-local theory of turbulence).
To obtain turbulent velocities $v_\mathrm{conv}$ from the convective enthalpy flux, one exploits that ${v}''^2\sim {h}''$  often holds between the turbulent velocity and enthalpy perturbations so that we can estimate the convective velocity by factorising, 
\begin{equation}
F_\mathrm{conv}^\mathrm{h} = 4\pi r^2 \langle\rho {v}'' {h}'' \rangle \approx
4\pi r^2\rho \langle {v}''^2\rangle^{1/2} \langle {v}''^2\rangle
\approx
4\pi r^2 \rho\langle {v}''^2\rangle^{3/2}.
\end{equation}
By setting $v_\mathrm{conv}=\langle v''^2\rangle^{1/2}$ and possibly also assuming that the convective energy flux $F^\mathrm{h}_\mathrm{conv}$ can be replaced with the total neutrino energy flux $\sim {F}_\mathrm{rad}^\nu$, one obtains the approximate convective velocity.
\begin{align}
    v_\mathrm{conv}^\mathrm{appr} &\approx
    \left(\frac{{F}^h_\mathrm{conv}}{4\pi r^2\rho}\right)^{1/3}
    \approx
    \left(\frac{{F}^\nu_\mathrm{rad}}{4\pi r^2\rho}\right)^{1/3},\label{eq:v_conv approx}
\end{align}
as mentioned before.
Is it only the assumption $F_\mathrm{conv} \sim {F}_\mathrm{rad}^\nu$ that is violated for PCS convection, or is the usual assumption
$v''^2\sim h''$ about the velocity and enthalpy fluctuations also problematic?
Figure~\ref{fig:v_conv} compares the convective velocities and the self-consistent 3D RMS velocities at 260 ms post-bounce time. The convective velocities $v_\mathrm{conv}^\mathrm{appr}$ estimated from Equation~(\ref{eq:v_conv approx}) are shown as {solid} lines (3D: red, 2D: grey), root mean squared (RMS) velocities
$ v_\mathrm{conv}\equiv 
\sqrt{\langle {v_r''}^2 + {v_\theta''}^2+{v_\varphi''}^2\rangle}$
are shown as {dashed} lines}.\footnote{One might also perform this comparison for the radial velocity fluctuations alone, but since there is usually rough equipartition between the radial and non-radial turbulent kinetic energies on average, this just amounts to a rescaling
of the proportionality factor between $v_\mathrm{conv}$ and the RMS enthalpy fluctuations.} 
Due to the high ratio of the neutrino energy flux to the convective energy flux shown in Figure~\ref{fig:flux3}, {$v_\mathrm{conv}^\mathrm{appr}$} (defined in Equation ~\ref{eq:v_conv approx}) is clearly overestimating the {the ``real''}
$v_\mathrm{conv}$. Figure~\ref{fig:v_conv} shows, however, that the assumption $F_\mathrm{conv} \sim {F}_\mathrm{rad}^\nu$ is no the only issue. Comparing $\tilde{h}_\mathrm{RMS}$ and $v_\mathrm{conv}^2$ in 2D (grey) and 3D (red), we can see that $v_\mathrm{conv}^2 \ll \tilde{h}''$. The squared convective velocity fluctuations are a factor $4\texttt{-}6$ smaller compared to the RMS enthalpy fluctuations. This means that the estimated convective velocity must be corrected downward
by another factor $2\texttt{-}2.5$ if ${F}_\mathrm{conv}^h$ is known.

\begin{figure*}
\hfill
\subfigure[]{\includegraphics[width=7cm]{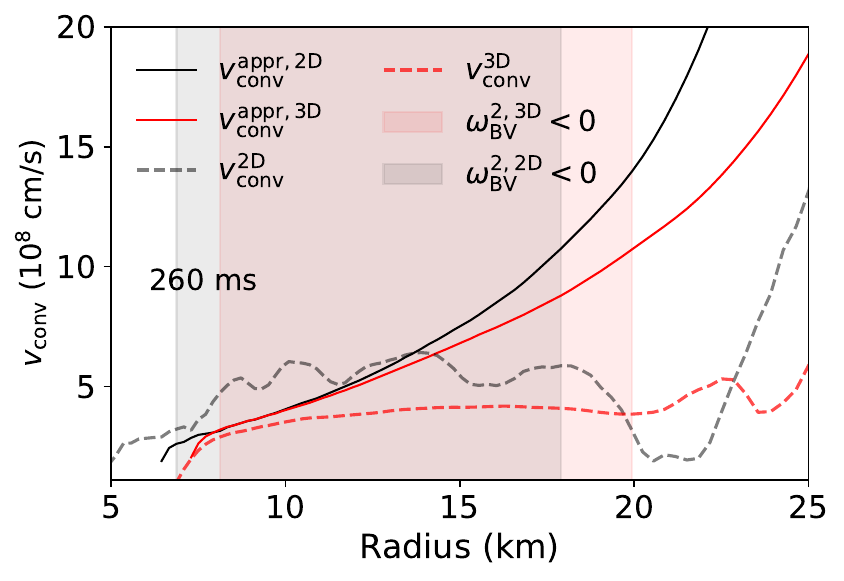}}
\hfill
\subfigure[]{\includegraphics[width=7cm]{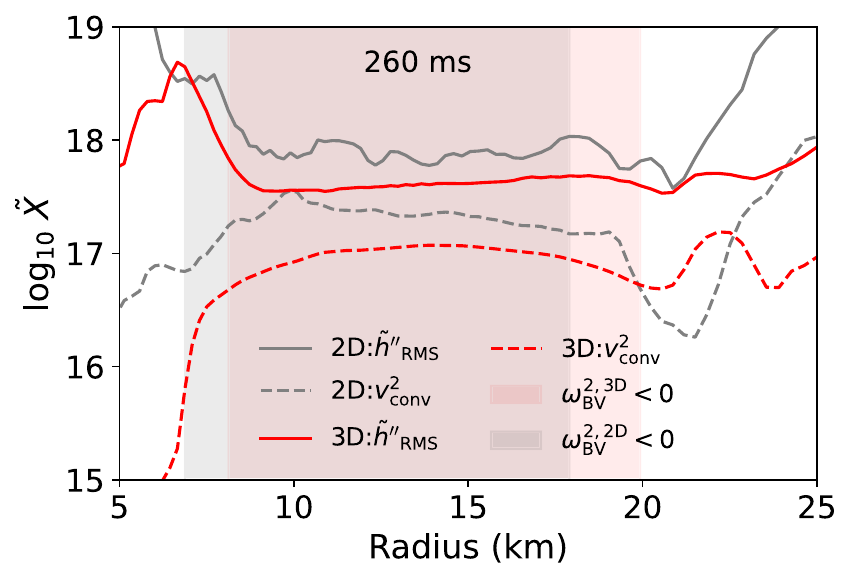}}
\hfill
\caption{(a) Convective RMS velocities  $v^\mathrm{2D/3D}_\mathrm{conv}$
in 2D {(dashed black)} and 3D {(dashed red)}  evaluated at $260\,\mathrm{ms}$ after bounce (dashed lines). In comparison to the  convective velocities $v^\mathrm{appr}_\mathrm{conv}$ estimated from Equqation~(\ref{eq:v_conv approx}) (solid lines). The unstable regions are shaded in red for 3D, and shaded in grey for 2D. (b) RMS enthalpy fluctuations $\tilde{h}''_\mathrm{RMS}$ (solid), and squared RMS velocity $v_\mathrm{conv}^2$ (dashed ) for 2D (grey), and 3D (red) at $260\,\mathrm{ms}$ after bounce. Analogously to (a), we show convective regions as a shaded background.}\label{fig:v_conv}
\end{figure*}

\subsection{Lepton number fluxes in the PCS convection zone}\label{sec:nuflux}
We next consider the transport of lepton number by convection in a manner analogous to the previous section.
Similarly to the energy transport, one expects that some form of balance condition determines the convective and diffusive lepton number fluxes. As we shall see, the balance between the convective and diffusive flux takes on a simpler form than for the energy transport.

%and investigate the interplay of stabilising and destabilising entropy and electron fraction gradients in quasi-steady convection, following similar lines as the earlier studies of \citet{powell_2019,glas_19}.

\begin{figure*}
	\centering	\includegraphics[width=0.8\linewidth]{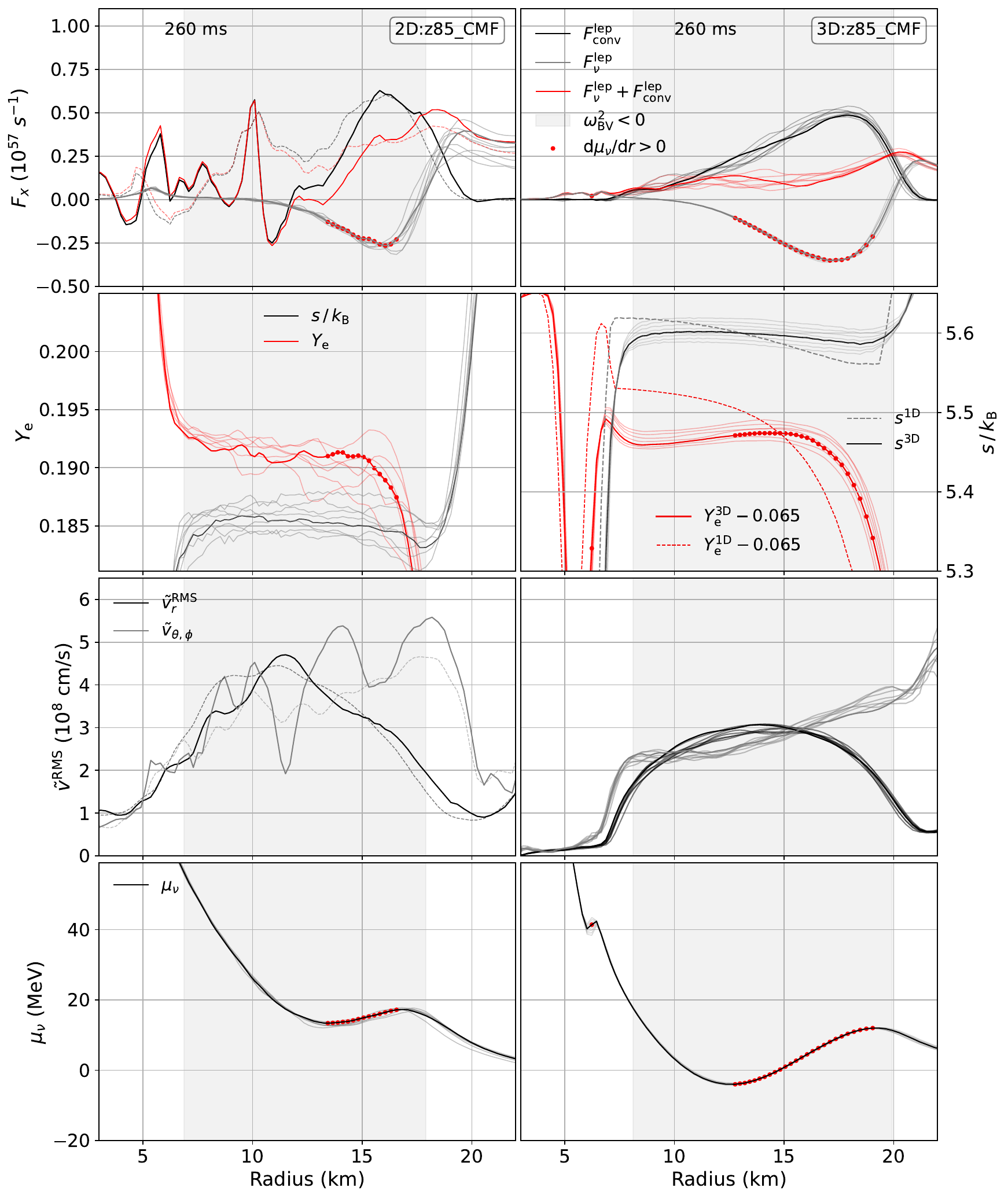}
\caption{Radial profiles for \texttt{z85:CMF} in 2D (left column) and 3D (right) at a post-bounce time of 260 ms. Row 1: Convective lepton number flux $F_\mathrm{conv}^\mathrm{lep}$ (black), diffusive net lepton number flux  $F_\nu^\mathrm{lep}$ (grey), and the sum of convective lepton number flux and diffusive net lepton number flux $F_\mathrm{conv}^\mathrm{lep}+F_\mathrm{\nu}^\mathrm{lep}$ (red). In 2D, dashed lines show averages over $\mathord{\sim} 12\,\mathrm{ms}$ and the light red and grey curves represent different time steps around $260\,\mathrm{ms}$ within a window $\mathord{\sim} 12\,\mathrm{ms}$. Row 2: Favre averaged entropy per baryon $\tilde{s}$ (black) and electron fraction $\tilde{\ye}$ (red, shown with an offset in 3D). Thin lines represent multiple timesteps of $\pm 2\,\mathrm{ms}$ around 260 ms. We also show the $\ye$ and $s$-profiles of a spherically symmetric simulation, denoted by 1D (dashed lines).
Row 3: RMS averages of radial velocity fluctuation $\tilde{v}_r^\mathrm{RMS}$ (black) and non-radial (grey) velocities evaluated as $\tilde{v}_{\theta,\phi}=\langle\sqrt{v_\theta^2 + v_\phi^2}\rangle$  
Row 4: Favre averaged neutrino chemical potential (black). Grey-shaded areas represent the regions of convective instability according to the Ledoux criterion. Red dots on lines in rows 1, 2 and 5 indicate the region where
the neutrino chemical potential has a positive gradient.
}\label{fig:flux1}
\end{figure*}

The first panel of Figure \ref{fig:flux1} displays the convective lepton number flux and the diffusive net lepton number flux from Equations~(\ref{eq:leptonflux}) and (\ref{eq:diffusive_net_lepton}) in 2D and 3D.
\begin{figure}
	\centering	
 \includegraphics[width=0.7\linewidth]{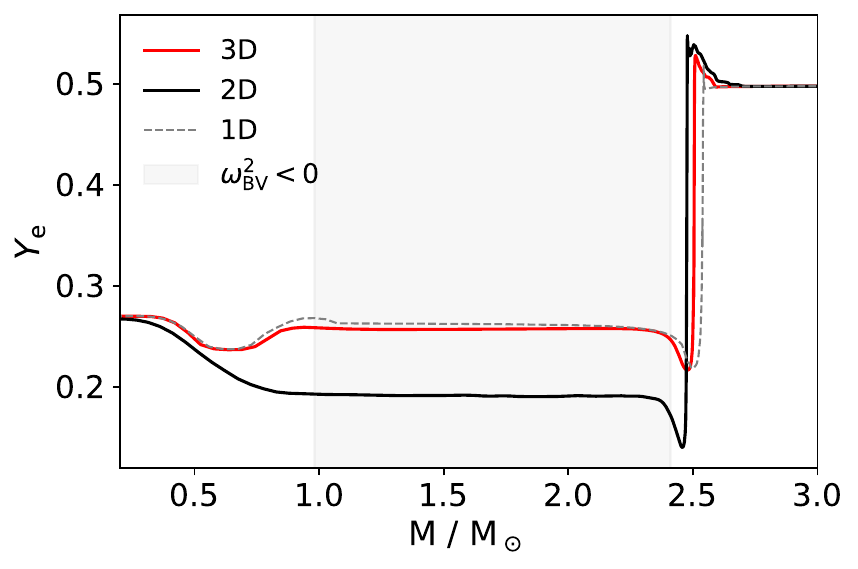}
\caption{Favre averaged electron fraction as a function of mass coordinate in 1D (grey dashed), 2D (black), and 3D (red) at a post-bounce time of $250\,\mathrm{ms}$. The electron fraction in the 3D model is significantly larger than in 2D.}\label{fig:ye_mass}
\end{figure}
The convective lepton number flux $F_\mathrm{conv}^\mathrm{lep}$ has a positive sign, whereas the net lepton number flux is generally negative during this phase~\citep{Pons1999-gs}. In 3D, both fluxes oppose each other at the outer edge of the PCS convection zone, and the net flux of diffusing and convective lepton number is almost constant, which was also discussed in \citet{powell_2019}.
Thus, a simple balance of fluxes approximately holds for lepton number transport in the PCS convection zone, contrary to the case of energy transport. This can again be understood as a result of self-adjustment of the fluxes in quasi-steady state.
A strong loss of net lepton number at the convective boundary will steepen the $\ye$ profile, which increases the convective lepton number flux $F^\mathrm{lep}_\mathrm{conv}$ to restore the balance. Conversely, if ${F}^\nu_\mathrm{lep}$ is small, the $\ye$ gradient becomes flatter, which stabilizes against convection, and the convective lepton flux consequently decreases. This self-adjustment leads to a quasi-equilibrium between the outward-directed convective lepton
number flux and the inward-directed diffusive lepton number flux, {so that ${F}_\mathrm{conv}^\mathrm{lep} + {F}^\nu_\mathrm{lep}\approx \mathrm{const.}$.}

The key difference to the energy transport is the high optical depth for electron neutrinos and antineutrinos. This leads to a rather small net lepton flux from the outer edge of the PCS convection zone, whereas the net energy flux in the PCS convection zone is largest at its outer boundary.

In this situation, the convective and diffusive lepton number flux end up balancing each other almost exactly to ensure roughly uniform deleptonisation throughout the convection zone.

\begin{align}
\dot{Y}_\mathrm{e,conv}+\dot{Y}_{\mathrm{e},\nu} & \approx 
\label{eq:balance}
-\frac{m_\mathrm{B}}{4\pi \rho r^2 \phi^6}
\frac{\partial}{\partial r} \left[\alpha \phi^4\left(
 F_\mathrm{conv}^\mathrm{h}+
F_\mathrm{rad}^\nu\right)\right] 
\end{align}

In 2D, the behaviour does not quite conform to this pattern as shown by Figure~\ref{fig:flux1}, but some form of balancing behavior of both fluxes may also be visible towards the outer convective boundary, where the {sum of convective lepton number flux and diffusive lepton number flux at the outer convective boundary are roughly constant (corresponding to small values of the sum of the divergences)}.
The negative diffusive lepton number flux is {mainly}
due to a positive gradient in the neutrino chemical potential $\mu_\nu$;
 {but note that the region of a negative diffusive lepton
 number flux extends beyond the region with $\partial \mu_\nu/\partial r>0$.}\footnote{In general, the diffusive lepton number flux is determined by the profile of $\mu_\nu$ and temperature and by
 various transport coefficients for electron neutrinos and antineutrinos,
 which can result in an inward flux even when $\partial \mu_\nu/\partial r<0$; see Equation~(30) of \citet{Pons1999-gs}.}
In 3D, we see $\D\mu_{\nu}/\D r > 0$ in between $12.5\,\mathrm{km}\lesssim r \lesssim 19\,\mathrm{km}$ (see red dotted points in the top and bottom row of Figure~\ref{fig:flux1}). 

\subsection{Interplay of Stabilising and Destabilising Gradients in PCS Convection}
The balance conditions discussed in Sections~\ref{sec:conv_velocities} and
\ref{sec:nuflux} regulate the convective energy and lepton number fluxes during steady-state convection such that a secular build-up of unstable gradients is avoided. The actual stratification of the PCS convection zone needs to be marginally unstable to sustain convective overturn, however. {It is worth investigating} the interplay of entropy and lepton number gradients in maintaining marginal instability more closely, extending similar analyses of instability in PCS by
\citet{Buras2006-nt,glas_19,powell_2019}.
%The negative gradient of $\mu_\nu$ also has implications for
%convective stability and instability \citep{powell_2019}.

Instability against convection in the adiabatic regime (i.e., neglecting diffusive effects) is governed by the Ledoux criterion for entropy and electron fraction gradients \citep{Ledoux_1947,Buras_2006a},
\begin{align}
 C_\mathrm{Ledoux} = \left(\pdv{\rho}{s}\right)_{\ye,P} \dv{s}{r} + \left(\pdv{\rho}{\ye}\right)_{P,s} \dv{\ye}{r} > 0.
 \label{eq:cledoux_s_ye}
\end{align}
The pre-factor $\left(\partial{\rho}/{\partial s}\right)_{\ye,P}$ is generally negative, so that a negative entropy gradient acts as destabilising.
The second thermodynamic derivative $\left(\partial{\rho}/{\partial\ye}\right)_{P,s}$ can have either sign, so that there is no straightforward stability criterion for the sign of $\D\ye/\D r$. 
\citet{powell_2019} showed that a negative lepton number gradient $\D\ye/\D r$ acts as stabilising  according to the Ledoux criterion 
in the presence of a positive neutrino chemical potential gradient since the Ledoux criterion can be rewritten as (see Appendix~\ref{subsec:mr}),\footnote{There is a typo in \citet{powell_2019}; the sign in their Equation~(\ref{eq:chempot_grad}), in front of $\rho^2$ should be negative, instead of positive. For a detailed calculation, see Appendix~\ref{subsec:mr}.}
\begin{align}
{
   \left(\pdv{\rho}{s}\right)_{\ye,P}
   \dv{s}{r}}
   - \rho^2
   \left(\pdv{\mu_\nu}{\rho}\right)_{s,\ye}
   \dv{\ye}{r}
   &> 0.\label{eq:chempot_grad} 
\end{align}
%where $\mathrm{d}s/\mathrm{d}r$ applies to good approximation in a well-mixed convection zone.
%Note, however, that the opposite is \textit{not} true for a negative radial neutrino chemical potential gradient.
\citet{powell_2019} suggested approximating the thermodynamic
derivative
$(\partial \mu_\nu/\partial \rho)_{s,\ye}$ by the
actual gradient $\ud \mu_\nu/\ud\rho$ to roughly determine
whether positive or negative $\ye$-gradients are conducive to instability. Upon evaluating the thermodynamic derivative
exactly, it turns out, however,
that $(\partial \rho/\partial \ye)_{P,s}>0$ \emph{throughout} the convective zone, even regardless of the sign of
$\ud \mu_\nu/\ud\rho$.
{Thus, positive (negative) $\ye$-gradients are always de-stabilising (stabilising) in the PCS convection zone in our models.}

The instability criteria (\ref{eq:cledoux_s_ye},\ref{eq:chempot_grad})
lead to a complex interplay of stabilising and destabilising entropy and electron fraction gradients in the PCS convection zone.

\begin{figure}
	\centering	
 \includegraphics[width=0.9\linewidth]{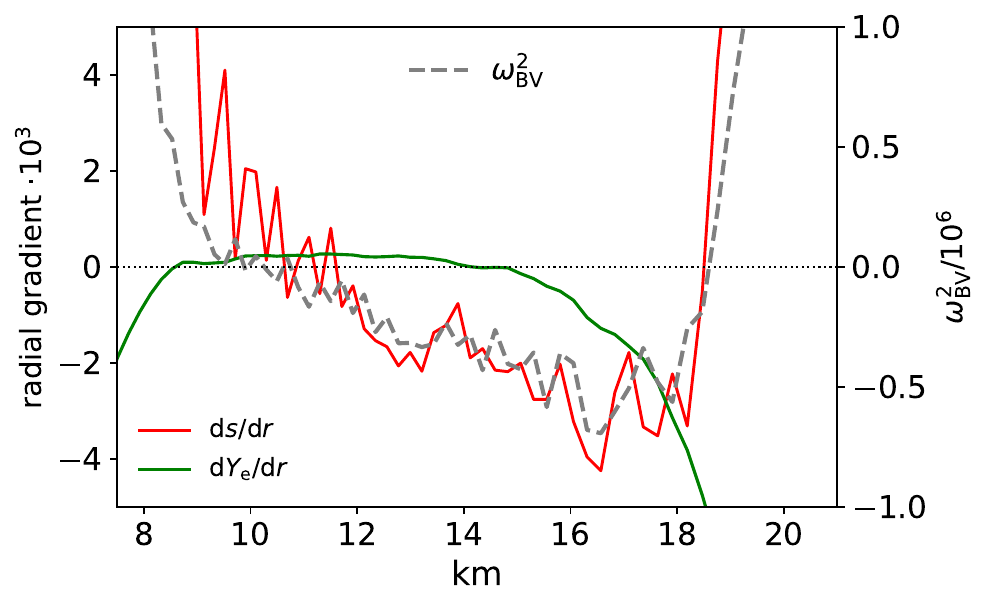}
\caption{{Favre-averaged radial electron fraction and entropy gradients (left y-axis), and \bv (right y-axis), both a function of radius for our 3D run. As the thermodynamic derivatives in Equation~(\ref{eq:cledoux_s_ye}) each carry a constant sign throughout the convective region, the Ledoux criterion for stability only depends on the sign of the averaged radial gradients. The Brunt-V\"ais\"al\"a frequency has a very similar radial
dependence as the entropy gradient.
In the region outside $14\,\mathrm{km}$, where the electron fraction gradient becomes stabilising, the Brunt-V\"ais\"al\"a frequency remains negative. Thus, the radial entropy gradient clearly dominates convective instability in most of the convective region. See text for a detailed explanation. }}\label{fig:ledoux}
\end{figure}

{To further illustrate, we show averaged radial entropy and electron-fraction gradients in Fig. \ref{fig:ledoux}. }
{We find that the entropy gradient remains positive in the outermost and innermost parts of the convection zone, thus, following the definition of the Ledoux criterion in Equation \ref{eq:cledoux_s_ye}, acting \textit{stabilizing} as the prefactor $\left(\partial \rho/\partial s\right)_{\ye,P}$ is generally \textit{negative}. However, in most of the convective region it is negative so that we find destabiling behaviour here. 
Regarding the electron fraction gradient, we discussed previously that the pre-factor $\left(\partial\rho/\partial\ye\right)_{P,s}$ is generally \textit{positive} (in our model) so that a positive $\ye$-gradient acts as destabilising and a negative $\ye$-gradient as stabilizing: Between $8\,\mathrm{km}\mathord{\lesssim} R\mathord{\lesssim{14}}\,\mathrm{km}$, $\D\ye/\D r$ is positive (\textit{destabilising}); in between $14\,\mathrm{km}\mathord{\lesssim} R\mathord{\lesssim{20}}\,\mathrm{km}$, it becomes increasingly negative (\textit{stabilising}). However, in most regions, the $\ye$-gradient plays a less significant role regarding overall convective stability compared to the entropy gradient. This trend of a dominating entropy fraction gradient is also evident by comparing the shapes of the entropy fraction gradient and the \bvns, which are orders of magnitude distinct, yet their shape is rather similar, indicating that the pre-factor, i.e., thermodynamic derivative $\left(\partial{\rho}/\partial{s}\right)_{Y_\mathrm{e},P}$ does not change substantially in this region (otherwise the shape of the entropy fraction gradient would not align anymore so well with the \bvns). These findings align with previous considerations in \cite{Jakobus_2023}, where the entropy term in Equation~(\ref{eq:cledoux_s_ye}) (including its prefactor) was dominant too and dictated the shape of the \bvns. Towards the outer PCS convective region, both gradients counter-act each other with $\D\ye/\D r < 0$ (destabilising) and $\D s/\D r > 0$ (stabilising). }
%There is a narrow region closer to the center of the convection zone, between about 15 km and 18 km, where both entropy, and electron-fraction gradient act as destabilizing. Just a bit further out, the negative entropy gradient becomes even more pronounced due to neutrino cooling, further increasing the destabilizing effect. However, this is counterbalanced and still dominated by a positive (and thus now \textit{stabilizing}) electron-fraction gradient in this outer region.}

As a side note, we see a significantly higher electron fraction in 3D than in 2D. The axes for the electron fraction in 2D and 3D carry different scales. This is better visible in Figure~\ref{fig:ye_mass}, where we plot $\ye$ as a function of mass. Compared to 2D, the electron fraction in 3D is increased by about $50\%$.
This surprisingly large difference is due to the extra release of lepton number in 2D during
the phase of prompt convection, which can be quite violent in 2D and magnifies the aforementioned differences in the structure of the core after bounce from slightly different collapse dynamics (Section~\ref{explosion_dynamics}).

\begin{figure*}
	\centering	\includegraphics[width=0.8\linewidth]{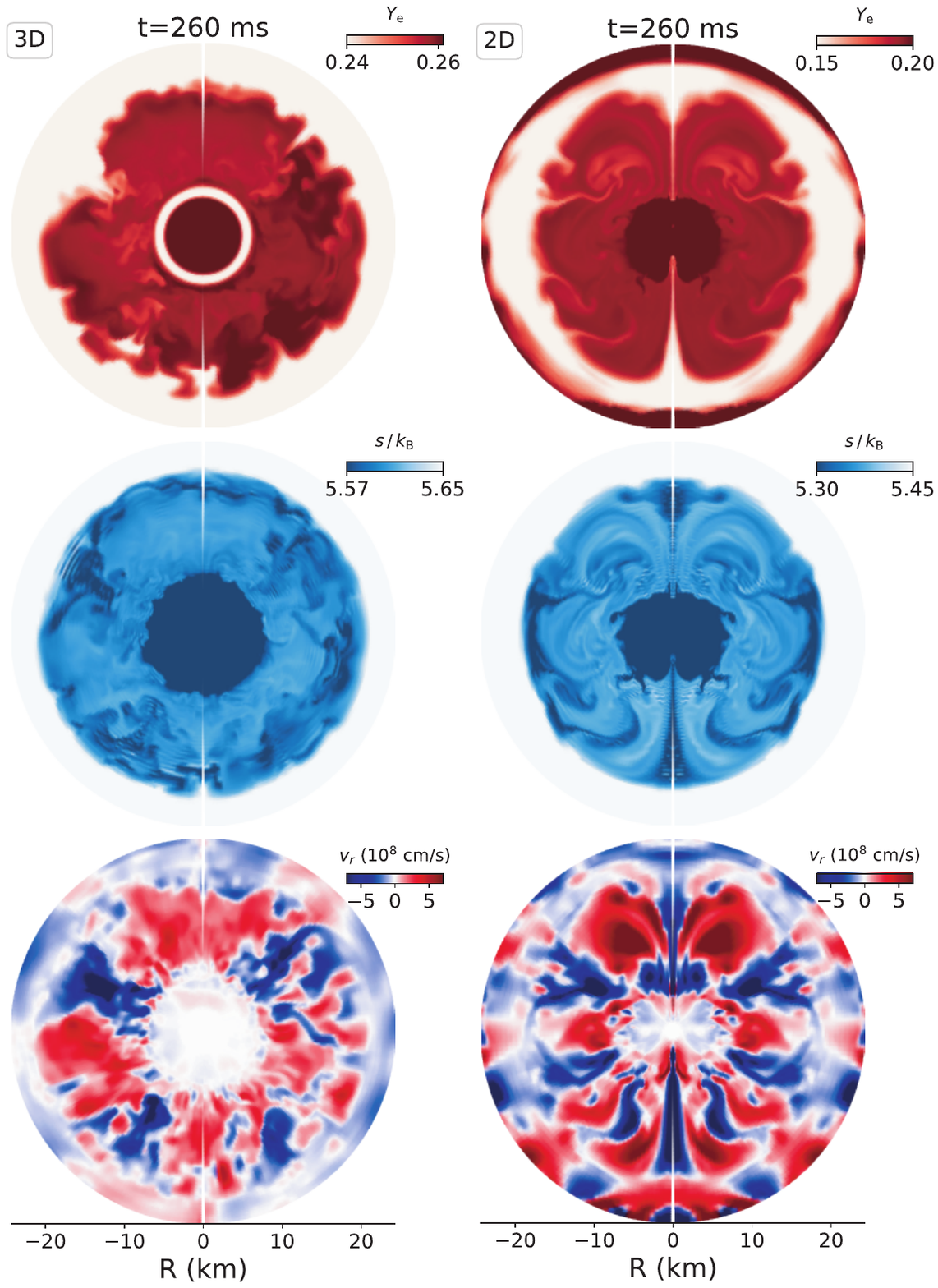}
\caption{Slices along meridional plane showing the electron fraction $\ye$ (first row), entropy per baryon $s$ (second row), and radial velocity $v_r$ (third row) for 3D (left column), and 2D (right column). The PCS convection zone stretches from approximately 20 km to 8 km (visible as enhanced radial velocities in the third panel). The electron fraction in the 3D run shows a small asymmetry on the lower right side (higher $\ye$). In 2D, and to a lesser extent in 3D, we see Ledoux convection, as low entropy and $\ye$ material sinks inwards. The structure in 2D is more distributed towards large scales, as expected from an inverse cascade.}\label{fig:radial_conv1}
\end{figure*}

\subsection{Comparison to Mixing Length Theory}\label{sec:mlt}
In order to systematically explore the dependence of the neutrino signal \citep{roberts_12,Mirizzi_2016} and the frequency trajectory of PCS oscillations modes \citep{wolfe_23}, e.g., on the nuclear EoS or progenitor mass and metallicity at manageable computational cost, 1D simulations remain indispensable. Especially for predicting synthetic mode frequency trajectories, the modification of PCS stratification by multi-dimensional effects must be taken into account. For this purpose, one can resort to mixing-length theory (MLT, as implemented, e.g., in \citealp{roberts_12,Mirizzi_2016}) or generalisations thereof \citep{mueller_2019}.
In MLT, turbulent processes are modelled as a diffusive process, and the convective flux is carried by one-scale eddies whose dimension is of the order of the \textit{local} pressure scale height (introduced below)~\citep{Prandtl_1952,Vitense_1958,weiss_2004}.
It is useful to check how well the assumptions inherent in MLT hold in the PCS convection zone and to what extent MLT can reproduce the stratification of 3D models. For this purpose, we have conducted a 1D simulation of model \texttt{z85:CMF} with the mixing-length module in \textsc{CoCoNuT-FMT}, and we also compare turbulent fluxes and fluctuations from a Favre decomposition of the 3D model to MLT theory.

Before such a comparison, it is useful to review some basic ingredients of MLT.
The turbulent diffusivity in MLT is given by $\Lambda_\mathrm{mix} {v}''_\mathrm{mix}$. The mixing length parameter $\Lambda_\mathrm{mix}$ is typically set as a fixed multiple of the pressure scale height,
\begin{align} 
H_\mathrm{P}=  -\tilde{P} \left(\dv{\tilde{P}}{r}\right)^{-1}=\frac{\tilde{\rho}\,g}{\tilde{P}},
\end{align}
where Favre averages (denoted by tildes) or Reynolds averages (denoted by hats or $\langle .\rangle$) need to be used in case ``virtual'' MLT fluxes are to be evaluated for a 3D model.
The MLT convective velocity is given in terms of $\Lambda_\mathrm{mix}$, the \bv from 
Equation~(\ref{eq:bv1}), and a dimensionless parameter $\alpha_1$ as
$v_\mathrm{conv}
= \alpha_1 \Lambda_\mathrm{mix}\omega_\mathrm{BV}.$
The MLT lepton number flux is then computed as
\begin{align}
    \langle\rho {v}''{\ye}''\rangle 
    &\approx \alpha_1\alpha_2\hat{\rho}v_\mathrm{conv}\Lambda_\mathrm{mix} \dv{\tilde{\ye}}{r}
    =\alpha_1\alpha_2\hat{\rho}\omegabv\Lambda_\mathrm{mix}^2\dv{\tilde{\ye}}{r},
\end{align}
where $\alpha_2$ is a second dimensionless coefficient. The convective energy flux in MLT is given by 
\begin{align}
    \langle\rho{v}''{h}''\rangle
    &\approx
    -\alpha_1\alpha_3\hat{\rho}v_\mathrm{conv}\Lambda_\mathrm{mix}\left(\dv{\tilde{\epsilon}}{r} + \tilde{P} \dv{\hat{\rho}^{-1}}{r}\right)
    \nonumber
    \\
    &=-\alpha_1\alpha_3\hat{\rho}\omega_\mathrm{BV}\Lambda_\mathrm{mix}^2\left(\dv{\tilde{\epsilon}}{r} + \tilde{P} \dv{\hat{\rho}^{-1}}{r}\right),
\end{align}
Consistency with the second law of thermodynamics requires
$\alpha_2=\alpha_3$ \citep{mueller_2019}. In our analysis, we set all the dimensionless coefficients to one, $\alpha_1=\alpha_2=\alpha_3=1$.

\begin{figure*}
	\centering	\includegraphics[width=0.8\linewidth]{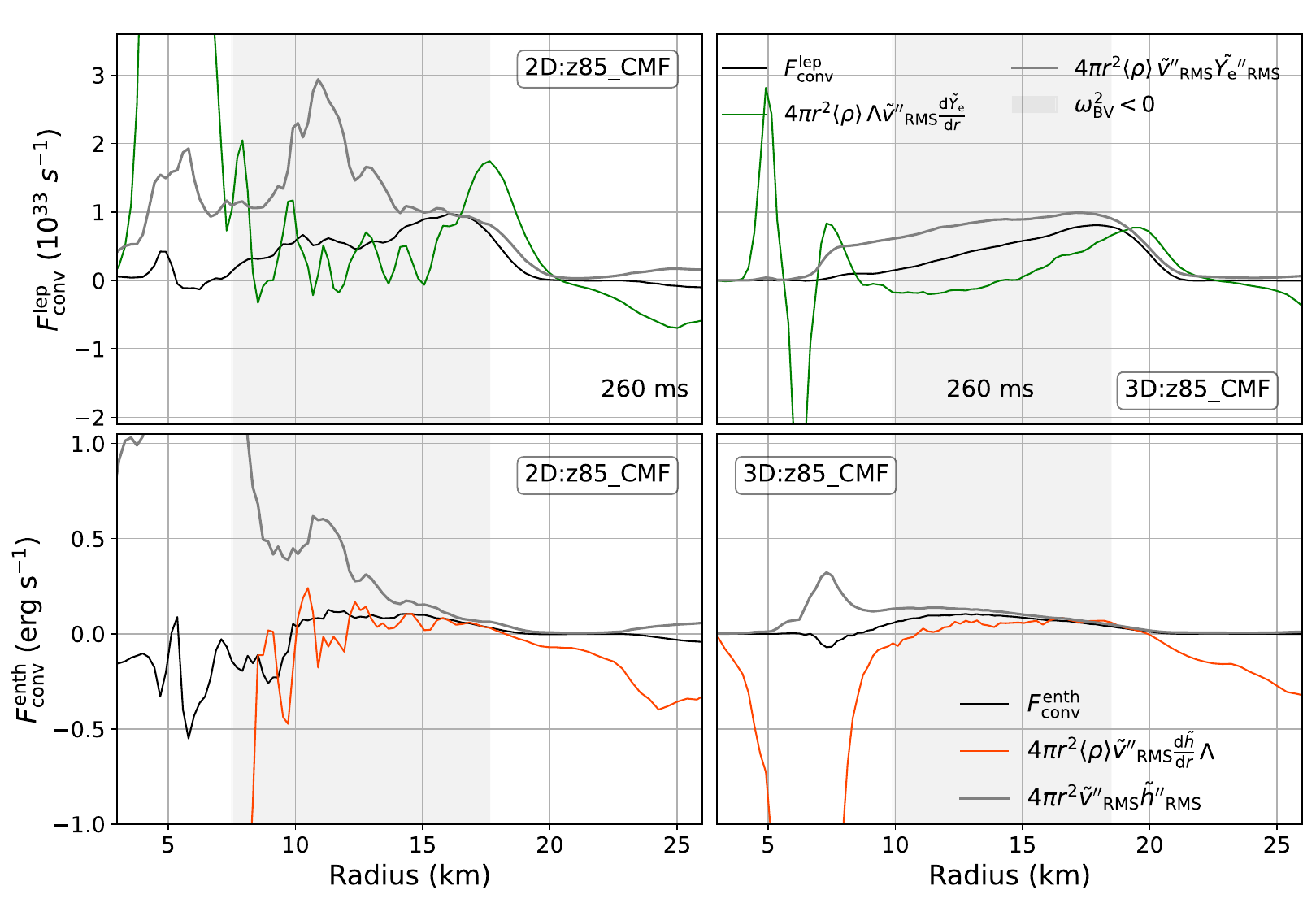}
\caption{Top row: Convective lepton flux $F_\mathrm{conv}^\mathrm{lep}$ (black), MLT gradient approximation (green) and factorised lepton flux (grey) for the 2D (left) and 3D (right) run. Bottom row: Convective energy flux $F_\mathrm{conv}^\mathrm{h}$, MLT gradient approximation (orange) and factorised energy flux (grey). The convective energy flux and MLT approximation for the energy flux in 2D are time-averaged over $15\,\mathrm{ms}$. For discussion, see Section~\ref{sec:mlt}.}\label{fig:flux2}
\end{figure*}

In Figure~\ref{fig:flux2}, we compare the MLT approximations to the actual convective lepton flux $F_\mathrm{conv}^\mathrm{lep}$ and the convective energy flux $F_\mathrm{conv}^\mathrm{h}$ from the 2D and 3D simulation.  In addition, we 
test the approximation of optimal (linear) correlation between the turbulent fluctuations that is implicit in MLT, i.e., the approximation that the Pearson coefficient of i) radial velocity and electron fraction fluctuation and ii) radial velocity and enthalpy fluctuations equals $\pm 1$.
To this end, we show factorised lepton 
 and energy fluxes, $4 \pi \langle\rho\rangle v''_\mathrm{RMS}
{Y_\mathrm{e}''}_\mathrm{RMS}$ and
$4 \pi \langle\rho\rangle v''_\mathrm{RMS} {h''}_\mathrm{RMS}$ (where $X''_\mathrm{RMS}$ denotes RMS fluctuations of any quantity). This allows us to to better recognise whether MLT is limited i) by the approximation of fluctuations through local gradients or ii) by the assumption of correlations between the radial velocity and thermodynamic variables.

Positive and negative fluxes correspond to upward and downward convective lepton number flows. Typically, $\ye$ is higher within the PCS core, so when high $\ye$ material is convectively transported outwards into regions with lower $\ye$, the flux $\langle\rho {v}''{\ye}''\rangle$ will be positive. Likewise, low $\ye$ material that is being transported inwards will result
in a positive (outward) flux of lepton number.
Both 2D and 3D simulations show a similar profile of the turbulent lepton number flux across the PCS convection zone, although the profile is less smooth in 2D.
The factorised flux $4 \pi \langle\rho\rangle v''_\mathrm{RMS}
{Y_\mathrm{e}''}_\mathrm{RMS}$ tracks the actual convective lepton number flux in 3D relatively well in the outer part of the convection zone, but overestimates the actual flux considerably near the inner boundary. This indicates less correlated velocity and electron fraction perturbations near the inner convective boundary. The MLT gradient approximation roughly captures the maximum convective lepton number flux, but even has the wrong sign in the inner part of the convection zone. The comparison of the actual convective lepton number flux is of less relevance in the artificial case of 2D axisymmetry. However, it is still included in the plots for the sake of completeness.

\begin{figure}
	\centering	\includegraphics[width=0.9\linewidth]{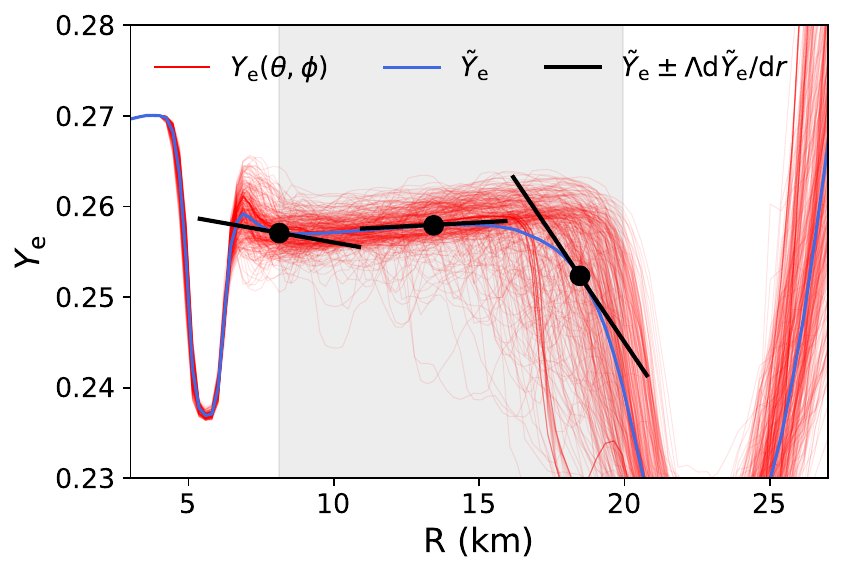}
\caption{Radial profiles of the electron fraction $\ye$ along different angular rays (red) and Favre averaged electron fraction $\tilde{\ye}$ (blue). The gradient of $\tilde{\ye}$ is drawn (black lines) for a mixing length ``step'' $\pm \Lambda \D\tilde{\ye}/\D r$  at three distinct radial positions (black dots) in the PCS convection zone (grey-shaded region). The slopes of the lines show the radial gradient; their extent in the radial direction is $\Lambda(r_i)$, evaluated as one pressure scale height, and their vertical extent gives the MLT density contrast. In the middle of the PCS convection zone, the flat gradient does not capture the dispersion of $\ye$ in rising and sinking bubbles; note in particular the region between $10\,\mathrm{km}\leq r\leq 15\,\mathrm{km}$.}\label{fig:mixinglength}
\end{figure}

The reason why the (gradient-based) MLT approximation is generally not very accurate and becomes increasingly worse towards the inner convective boundary is better understood from Figure~\ref{fig:mixinglength}. Displayed are radial profiles of $\ye$ for different angles (latitude and longitude). The Favre average $\tilde{\ye}$ is overlayed as a blue line, together with black lines indicating the gradients $\partial \tilde{\ye}/\partial r$ at three positions (black dot); in the middle and at the outer and inner boundary of the PCS convection zone. The radial projections of the black lines show the mixing length $|\pm\Lambda|$. Comparing the vertical extent of the gradients and the dispersion of the red curves shows how well the actual $\ye$-contrast in rising and sinking plumes is captured by the extrapolated gradient across one pressure scale height.
In regions where the angle-averaged profile is flat, such as in the centre of the convective boundary, the dispersion of $\ye$ is not well captured by the gradient approximation of MLT. The figure also reveals why MLT underestimates the actual convective lepton number flux and sometimes even predicts the wrong sign for the flux. 
The gradient of $\tilde{\ye}$ becomes positive, but is relatively small in the middle of the convection zone. Hence, MLT predicts a small dispersion of $\ye$, and
excess $\ye$ in downdrafts and a $\ye$ deficit in updrafts.
In reality, large eddies that travel beyond the mixing length transport material with very high or low $\ye$ into regions characterised by a flat local $\ye$ gradient. As a result, there can be a significant dispersion of material from relatively large rising and sinking plumes, \textit{despite} a flat angular averaged radial profile. Moreover, turnover motion over the entire convection zone will transport low-$\ye$ material from the outer boundary downward and high-$\ye$ material outwards, so that the correlation of velocity and $\ye$ fluctuations has a different sign than expected from the local gradient.
To capture these features, one may require a non-local theory of turbulence that better tracks the evolution of $\ye$ along the
pathlines of convective flow, appropriately averaged over eddies of different scales.

The significant overestimation of the actual lepton flux by the factorised flux $4 \pi \langle\rho\rangle v''_\mathrm{RMS}
{Y_\mathrm{e}''}_\mathrm{RMS}$ near the lower boundary can also be related to the actual spatial structure of the convective flow.
When large eddies fragment into small-scale turbulent motions upon reaching the convective boundary, the flow becomes dominated by random motions on small scales, so that ${v}''$ and ${\ye}''$ will become essentially uncorrelated.

In the second panel of Figure~\ref{fig:flux2}, we show convective energy fluxes for 2D (left) and 3D (right). The pattern is similar to what we observed for the lepton flux. The actual convective flux in 3D shows quite good agreement with the factored flux $4 \pi \langle\rho\rangle v''_\mathrm{RMS} {h''}_\mathrm{RMS}$
near the outer convective boundary and in the middle of the convection zone (indicating a good correlation of velocity and enthalpy in updrafts and downdrafts)
but deviates from the factored flux towards the inner convective boundary. 
Similarly to the convective lepton flux in the upper panel, the gradient approximation $h''=\Lambda_\mathrm{mix}\D\tilde{h}/\D r$ gives the wrong
sign for the MLT flux in parts of the convection zone around $\mathord{\sim}11\,\mathrm{km}$, although the MLT approximation appears to match somewhat better for energy transport than for lepton transport.

Despite the systematic errors in the MLT flux, the effect on the electron fraction and entropy profiles in dynamical simulation is limited. Comparing profiles from the 3D simulation and a 1D simulation with MLT in
Figures~\ref{fig:ye_mass} and \ref{fig:flux1}, we find that MLT captures the average level of entropy and $\ye$ in the PCS convection zone very well. However, the 1D MLT model ends up with consistently negative entropy and $\ye$ gradients in the convection zone. The variation of these quantities within the relatively well-mixed interior of the convection zone is more significant, although the profiles still remain quite flat. Such small deviations may not compromise accuracy very much in simulations of neutron star cooling, but they may be more relevant when computing GW eigenfrequencies based on 1D profiles, and accurate profiles of the \bv are required.

\subsection{Spatial structure of PCS convection}\label{sec:passive_scalar}
How could MLT be improved to better reproduce the correct convective energy and lepton number fluxes? For stellar convection, more general theories of convection have been proposed to incorporate non-local estimates of the contrast
in advected quantities and the existence of eddies of different scales (e.g., full-spectrum turbulence; \citealp{canuto_92,canuto_96}). In the most abstract terms, one might compute the contrast $\delta\ye\equiv\ye(r+\Lambda)-\ye(r-\Lambda)$ (or other fluctuations) non-locally instead of applying a first-order Taylor series, and average over a range of mixing lengths $\Lambda$ with some appropriate turbulence spectrum, perhaps after taking into account effects of the turbulent cascade and non-ideal effects on eddies of various sizes. This is far beyond the scope of this paper, but the concept of non-local mixing is a good starting point for analysing the spatial structure of PCS convection and revisit the electron fraction asymmetry in the PCS that is characteristic of the LESA phenomenon (lepton-number emission self-sustained asymmetry, \citealp{Tamborra_2014}) seen in many supernova simulations \citep[e.g.,][]{powell_2019,glas_19}.

To illustrate the multi-dimensional structure of the convective flow, we show 2D slices of the electron fraction $\ye$ (first row), the entropy per baryon~$s$ (second row), and the radial velocity $v_r$ (third row) for the 2D and 3D run at $260\,\mathrm{ms}$ after bounce in Figure~\ref{fig:radial_conv1} . In 3D and 2D, Ledoux convection pushes low entropy material with lower $\ye$ inwards, e.g., in the upper left (``11 o'clock'') of each panel both in 2D and 3D. In 3D, the patterns show more small-scale turbulence due to the forward turbulent cascade~\citep{Kolmogorov_1941}. High-$\ye$ material is asymmetrically ``bulked'' towards the lower right of the 3D run,
indicative of a large-scale LESA asymmetry.
In the 3D entropy plots, one can also see more clearly some sinking low-entropy plumes of smaller scale, e.g.,
a low entropy plume sinking inwards, e.g., at the 7 o'clock''
and 10 o'clock positions. Especially the plume at 10 o'clock is also visible as a high $\ye$ plume. Large-scale structures in radial velocity are also present in 3D as in \citet{glas_19},
but the small-scale structures in velocity appear more prominent
than in entropy and electron fraction. In 2D, the electron fraction, entropy, and radial velocity fields are all dominated by large-scale structures as a manifestation of the inverse turbulent cascade in 2D~\citep{Kraichnan_1967}. 

As expected, radial velocities in 2D are somewhat increased, and of larger scale, compared to 3D~\citep{Kraichnan_1967}.

\begin{figure*}
	\centering	\includegraphics[width=\linewidth]{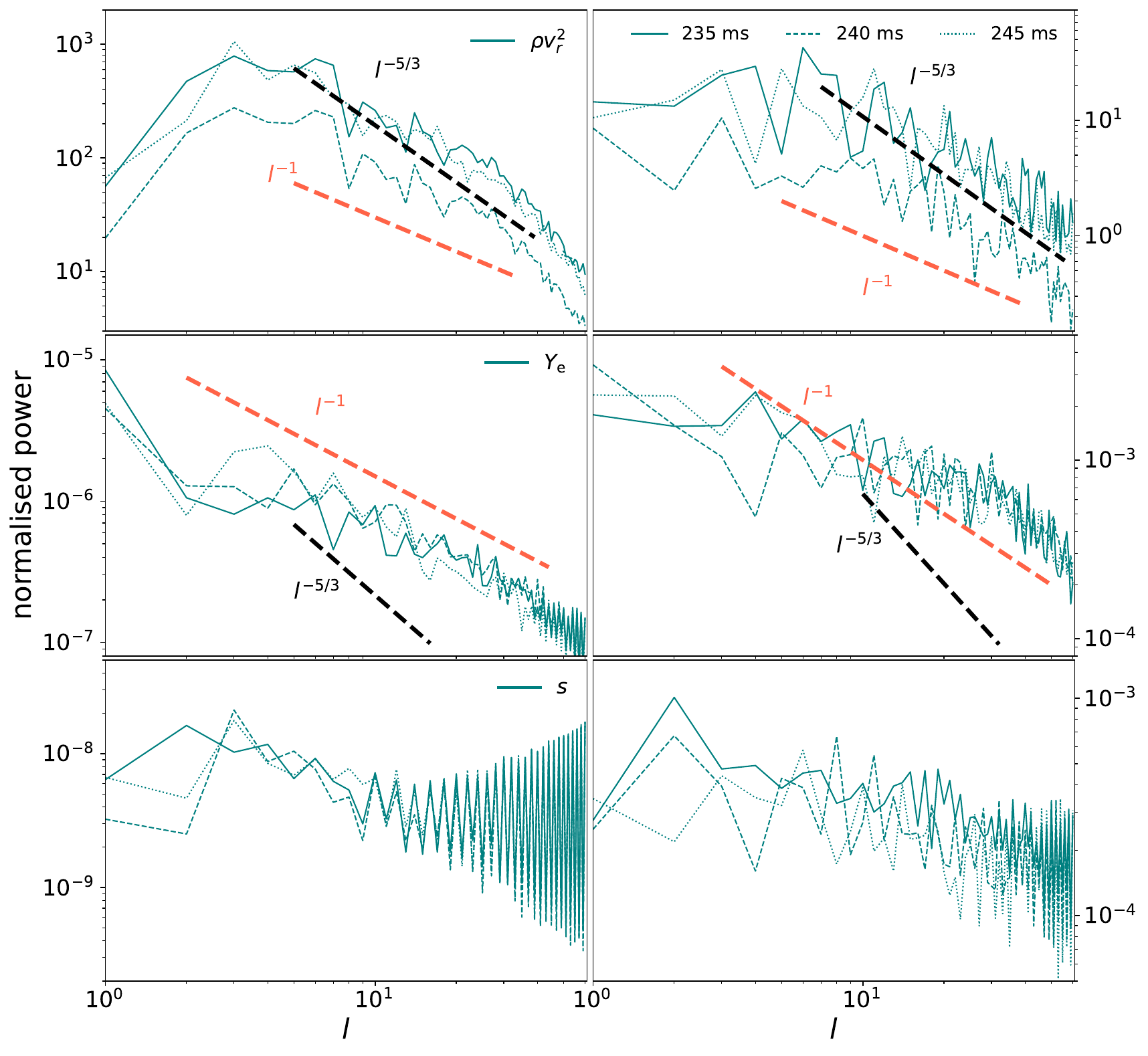}
\caption{Power spectra for $\rho v_r^2$, electron fraction $\ye$ (second row), and entropy per baryon $s$ (third row), computed according to Equation~(\ref{eq:passive_scalar}) in 3D (left), and 2D (right). The spectra are evaluated at $13\,\mathrm{km}$ for the 3D run, and at $11\,\mathrm{km}$ for the 2D run averaged over  the time interval $225\,\mathrm{ms}\leq t\leq 260\,\mathrm{ms}$ in both cases. All values have been normalised to the monopole $\hat{X}_0$. The radial velocity and electron fraction follow different scaling laws. The scaling law of the radial velocity follows a typical $5/3$ slope, whereas the electron fraction approximately follows a $l^{-1}$ slope. The power spectrum for the entropy is very noisy at large spherical harmonics degrees due to low-level odd-even patterns in the numerical solution. }\label{fig:passive_scalar}
\end{figure*}

To qualitatively understand the scale-dependent turbulent motions, we decompose the square roots of the radial Reynolds stress component $\rho v_r^2$, electron fraction $\ye$, and entropy $s$ per baryon into
spherical harmonics (Figure~\ref{fig:passive_scalar}). Spectra $\hat{X}_l$ of any quantity $X$ are calculated as,
\begin{align}\label{eq:passive_scalar}
   \hat{X}_l = \sum_{m=-l}^l\left|\int\D\Omega\mathcal{Y}_{lm}^*(\theta,\phi)\sqrt{X}\right|^2.
\end{align}
Taking the square root of these quantities ensures that the power in the decomposition of 
$\sqrt{\rho} v_r$ can be interpreted as the square root of the $rr$-component of the Reynold stress tensor (and half of the kinetic energy spectrum).
We normalise the  spectrum $\hat{X}_l$ by monopole $\hat{X}_0$. The procedure is similar to the decomposition of the turbulent kinetic energy in Equation~(\ref{eq:spectral}) (note that for the kinetic energy, we instead use the absolute of the radial velocity and omit a factor 0.5 in front of the sum). 

We observe a Kolmogorov scaling law for the velocity spectrum with a $5/3$-slope in the inertial range towards intermediate scales at angular wavenumbers above $l\gtrsim 12$.
We observe a flatter scaling law at lower polynomial degrees $l\lesssim 10$, with eventually decreasing power at low wavenumbers $l\lesssim 10$. 
\citet{powell_2019} already noticed the deviation from the Kolmogorov slope at lower wavenumbers and suggested that this may be related to less efficient driving of low-order modes due to the stabilising influence of the $\ye$-gradient in parts of the PCS convection zone. Still, it is difficult to test this hypothesis numerically. The entropy may exhibit a spectrum similar to $\ye$, but there is too much noise in the entropy spectrum already at low wavenumbers to diagnose this confidently.

The scaling looks different for the electron fraction $\ye$, which was also noted by \citet{powell_2019} as a yet unexplained feature.
 At low-intermediate wavenumbers $3 \lesssim l\lesssim 13$, the power spectrum approximately follows a $l^{-1}$ scaling law. As pointed out by \citet{powell_2019}, this disparity \emph{is} essentially LESA in the PCS, merely viewed in spectral space instead of real space. It reflects the presence of significant large-scale patterns in $\ye$ despite \emph{slow} overturn on large scales and relatively more power on medium scales in the velocity field.
 
 We propose that the different slope of the spectrum of $\ye$ is related to the 
 fact that the electron fraction is an advected scalar quantity (similar to a passive scalar, but slightly different because the electron fraction also influences buoyancy). Passive scalars typically follow a different scaling law in turbulent flows  \citep{batchelor_1959,SHRAIMAN_2000}. The distinct scaling arises from anomalous mixing (instead of the improbable event of an atypical path with typical mixing), specifically low probability configurations in which a fluid parcel travels from a large distance to the point of mixing without much mixing and dissipation along the way~\citep{Shraiman_1994}. The assumptions made by \cite{batchelor_1959}, where viscosity dominates over diffusion, do not apply to our case of an ideal hydrodynamic simulation with neutrino transport. However, the crucial point is that the scaling law of advected quantities will generally differ from the kinetic energy spectrum.
 
 This notion can be
 adapted to PCS convection and linked to the discussion about the limitations of MLT as a single-scale theory of convection.
If we retain the MLT notion that the $\ye$ contrast depends on the ``point of origin'' of material at a distance $\pm \Lambda$,
\begin{equation}
    \delta\ye\equiv\ye(r+\Lambda)-\ye(r-\Lambda),
\end{equation}
but treat $\Lambda$ as variable, representing the eddy scale; then one can motivate a Batchelor-type scaling law $\hat{Y}_\mathrm{e}\propto l^{-1}$.
As the angular wavenumber $l$ corresponds to the number of eddies that can be fit on a meridian between $\theta=0$ and $\theta=\pi$, the eddy scale is just $\Lambda = \pi R_\mathrm{conv}/l$
\citep{Foglizzo_2006}, where $R_\mathrm{conv}$ is the radius of the convective region (which we omit in the following as this is a ``constant'' parameter).\footnote{This assumes that the eddy is approximately symmetrical, that is, the horizontal extent of the eddy equals the vertical length, which we identify as the (radial) mixing length parameter $\Lambda$.}  Then, a Taylor expansion of
$\delta\ye\sim \ye(r+\Lambda)-\ye(r-\Lambda)$ immediately yields,
\begin{equation}
\label{eq:yespec}
    \delta Y_\mathrm{e}\propto \ye(r+\Lambda)-\ye(r-\Lambda) \propto l^{-1} \frac{\partial \tilde{Y}_e}{\partial r}.
\end{equation}
One should note that Equation~(\ref{eq:yespec}) is expected to break down for low $l$ because the gradient approximation is no longer appropriate and because the conversion from spatial scale to angular wavenumber becomes non-trivial (see the discussion on spherical Fourier-Bessel decomposition in \citealt{Ferndandez_2014}).

One can further rationalise this somewhat unintuitive scaling behaviour if we consider a balance of forcing and dissipation terms that 
create and destroy $\ye$-fluctuations
$\delta \ye (\Lambda)$ of a single-wavenumber mode on scale $\Lambda= \pi R_\mathrm{conv} l$,
namely generation from a
background gradient $\partial \ye/\partial r$ by velocity fluctuations and damping by turbulent dissipation with a dissipation time $\tau_\mathrm{diss}=\Lambda^2/D$
set by turbulent diffusivity on scale $D=\delta v_r (\Lambda) \Lambda $.
In such a simple model, the rate of
change of electron fraction perturbations on
scale $\Lambda$, $\delta \dot{Y}_\mathrm{e}$, is given by
\begin{equation}
\label{eq:dynamical_turbulence}
   \delta \dot{Y}_\mathrm{e}  (\Lambda)
    \approx \delta v_r(\Lambda) \frac{\partial \ye}{\partial r}
    -  \frac{\delta Y_\mathrm{e}}{\tau_\mathrm{diss}(\Lambda)}
    \approx
    \delta v_r(\Lambda) \frac{\partial \ye}{\partial r}
    -  \delta Y_\mathrm{e} \frac{\delta v_r (\Lambda) \Lambda}{\Lambda^2}.
\end{equation}
This leads to $\delta \ye (\Lambda) \propto  \Lambda \propto \pi/l$ in steady state ($\delta \dot Y_\mathrm{e}=0$), and suggests an $l^{-1}$ power spectrum. Equation~(\ref{eq:dynamical_turbulence}) neglects the cascade
or ``transfer'' terms that would appear in a full
spectral turbulence model
\citep[e.g.,][]{canuto_96} and replaces them with
a simple, scale-dependent eddy viscosity. Specifically, the transfer of power from $l'<l$
perturbations in $\ye$ to wavenumber $l$ are neglected as a generation term for fluctuations. Neglecting this contribution in favour of ``local'' generation of $\ye$ fluctuations at wavenumber $l$ can be justified by noting that compared to normal Kolmogorov turbulence, there is less power in velocity perturbations at small wavenumber $l$ to drive this transfer.\footnote{The transfer of
power in $Y_\mathrm{e}$ fluctuations from low to high $l$ is driven by triad interactions resulting from the Fourier transform of the advection term and will hence be determined by the power in fluctuations
$Y_{\mathrm{e},\mathbf{k}'}$ and 
$v_{r,\mathbf{k}''}$ at some wavenumbers $\mathbf{k}'$
and $\mathbf{k}''$ that match
some target wavenumber $\mathbf{k}=\mathbf{k}'+\mathbf{k}''$.
} The key idea here is that since the
velocity fluctuations on scale $\Lambda$ regulate both the driving and the damping of $\ye$ fluctuations on this scale, the spectrum of $\ye$ fluctuations may be largely independent of the velocity spectrum. By such a mechanism, stronger $\ye$ fluctuations at larger scales (smaller $l$) may emerge even if the velocity spectrum does not have appreciable power at low $l$.

Undoubtedly, much further analysis will be needed to analyse the driving, damping and transfer of power in the fluctuation spectra and possibly explain the peculiar spectral properties of PCS convection rigorously from turbulence theory. Nonetheless, we believe it is useful to outline a possible mechanism for explaining the distinct spectra of $Y_\mathrm{e}$ and radial velocity and hence for LESA.

\section{Conclusions}\label{sec:concl}
We studied the dynamics of proto-compact star (PCS) convection and its impact on gravitational emission from general-relativistic core-collapse supernovae in 2D and 3D simulations.

In particular, we investigated the gravitational wave
signal from the ${}^2\!g_1$ mode in the PCS core, which has
recently been identified as a potential probe of high-density equation-of-state physics in the 2D
simulations of \citet{Jakobus_2023}, and has also been
seen in recent 3D simulations of \citet{Vartanyan_2023}.

Although GW emission from this mode is not expected to be universally present in core-collapse supernovae, it is
useful for relating its frequency to interpretable PCS parameters.
Even when the signal from the core $g$-mode is not present, such a fit
could also be used to describe the emission gap that is often seen
in predicted supernova GW spectrograms at the avoided crossing of the ${}^2\!g_1$ and the dominant, rising $f/g$-mode emission band.
To this end, we showed that the frequency of the ${}^2\!g_1$ mode closely tracks the peak value of the \bv at the
edge of the PCS core. We further developed a fit formula
to approximate the frequency in terms of the mass and radius
of the low-entropy PCS core, the neutron star mass, the
entropy gradient between the PCS core and mantle, and thermodynamic derivatives at the edge of the core.
Among these quantities, the core mass and the entropy gradient
are relatively universal during the early post-bounce phase. Our fit formula is given by 
\begin{equation}
   \omegabv = c_\mathrm{gr}
    c_\mathrm{eos}
    \sqrt{4 \pi \dv{s}{m}}.
\end{equation}
with the prefactors 
\begin{align} 
c_\mathrm{gr} &= \left[\D\alpha/\D r\cdot r^2\alpha/(h\phi^4)\right]^{1/2}\,\,\,\mathrm{and}\\
c_\mathrm{eos} &= \left(\partial P/\partial s\right)_{\rho,\ye}^{1/2}/c_\mathrm{s}
\end{align} 
We tested our fit formula on 2D models for two simulations
(\texttt{z35:CMF} and \texttt{z85:CMF})
using $35 M_\odot$ and $85 M_\odot$ progenitors with the CMF
EoS \citep{motornenko_2020} and a $85 M_\odot$ progenitor using the SFHx EoS \citep{Steiner_13} (the GW signal was absent in the SFHx EoS setup for the $35 M_\odot$ progenitor).
Using the \bv (multiplied by a calibration factor $c_\mathrm{calib}=0.55$) to track the mode frequency produces good results. However, given our choice for the calibration factor, the \bv \textit{underestimates} the SFHx core $g$-mode frequency at later times. 
The fit formula, using our approximations for the terms arising in the \bvns, tracks the actual mode
frequently reasonably well early on, but distinctly overestimates
the mode frequency for \texttt{z35:CMF} at later times. 

Errors in the fit stem from different sources, and the most notable among
them are related to the assumption of a fixed mass coordinate for
the core-mantle interface:
\begin{enumerate}
    \item The largest error source is the mass term $M_\mathrm{mode}$  in  Equation~(\ref{eq:bv_approx})
    that stems from the term $\D\alpha/\D r$ in the \bv.
    Interestingly, we can approximate the term
    $\D\alpha/\D r$ fairly well by a simple fit
    \textit{although} it contains the troublesome mass coordinate term $M_\mathrm{mode}$. The reason is that $\D\alpha/\D r$ contains the surface gravity $M_\mathrm{mode}/r_\mathrm{mode}^2$ which is \textit{less} sensitive to the location of the mode.
    In the final fit formula for the mode frequency, the
    factor $r_\mathrm{mode}^{-2}$ is cancelled by another term,
    and the bigger error of $M_\mathrm{mode}$ degrades the overall
    fit.
    \item The second largest error source stems from the entropy gradient approximation $\D s/\D m = \mathrm{const}$. Penetration of material into the core $g$-mode region at later times, and neutrino diffusion (``cooling'') together flatten the gradient $\D s/\D m$. 
    \item The EoS term $c_\mathrm{eos}$ in the \bv
    is strongly sensitive to the choice of mass coordinate. The fixed mass coordinate approximation in the EoS term yields an error of up to $\mathord{\sim} 15\%$, which marks the third largest error source in our approximation. 
    \item The error for our estimate of the effective relativistic acceleration $c_\mathrm{gr}$  in the \bv
    is contained below $10\%$. The error mainly arises from the derivative of the lapse function $\D\alpha/\D r$. Due to cancellations of errors in individual terms, $c_\mathrm{gr}$ is only mildly sensitive to the choice of the fixed mass coordinate.
\end{enumerate}
Although the mode fit becomes inaccurate at a later time, it seems to work well for the position of the avoided crossing and could be applied to concluding PCS properties from the frequency
of the gap.
Based on our analysis of the factors entering the
\bv, we can also answer why the frequency of the mode is decreasing: The parameter for the effective relativistic acceleration, $c_\mathrm{gr}$, decreases as a function of time (both when evaluated at a fixed mass shell or when tracking the peak of the \bvns), 
as its variation is mostly driven by the decrease of the 
lapse function $\alpha$.\footnote{By contrast, the PCS surface $f/g$ mode frequency (which increases as function of time) is proportional to the surface gravity $M_\mathrm{mode}/r_\mathrm{mode}^2$.} Furthermore, the angle-averaged speed of sound profile becomes slightly flatter over time, and $\D s/\D m$ decreases by neutrino diffusion, and mixing. Together, these effects cause the ${}^2\!g_1$ mode frequency to \textit{decrease} as function of time. 
{An improved analytical understanding of mode frequencies and the properties of convection may be instrumental in the detection of future galactic CCSNe, particularly in the era of third-generation GW detectors.  Further simulations are needed to study when and how strongly modes other than the \textit{f}/\textit{g}-mode or the SASI are excited and detectable in gravitational waves.}
\\\\
In part II of our paper, we aimed to better understand what mechanisms precisely determine the excitation of the core $g$-mode. 
We found that the turbulent kinetic energy displays similar magnitudes in 2D and 3D. A spherical harmonic decomposition of the turbulent kinetic energy shows that the contribution to the quadrupolar mode exhibits similar strengths in both 2D and 3D. The inverse cascade in 2D is less established at small spherical harmonic degrees $l\lesssim 5$, and 2D and 3D spectra show similar behaviours here.  

We computed the autocorrelation function of the quadrupolar contribution to the turbulent kinetic energy and noticed shorter correlation times in 2D, coupled with increased power at higher frequencies in the Fourier space. A longer correlation time suggests reduced vigorous motion on a large scale, possibly leading to decreased dispersion in the eddy velocities. We hypothesise that eddies in the 2D configuration undergo more "impulsive" forcing and that their frequency exhibits a greater overlap with the natural core $g$-mode frequency, allowing for more resonant excitation in 2D, as previously suggested by \citet{andresen_2017}.

Although observed in the oxygen shell in \citet{mueller_2016} for nuclear burning, a stable stratification in the PCS convection zone is generally \textit{not} maintained by an equilibrium of the total neutrino energy flux and convective energy flux, where $F_\mathrm{rad}^\nu\mathord{\sim} F_\mathrm{conv}^\mathrm{h}$.
Instead, convective energy transport is governed by a more complex balance between
entropy source and sink terms due to the divergences of the convective and diffusive energy fluxes and the divergences of the convective and diffusive lepton number fluxes, which results in roughly constant secular rates of entropy change across the PCS convection zone.
Consequently, convective velocities are lower than expected from the total neutrino energy flux. We further notice that the general assumption that $v''^2 \mathord{\sim} h''$ neither holds in the PCS convection zone. 
The situation is different for lepton number transport, governed by a simpler balance condition.

Upon analysing the radial profiles of electron fraction, we find that the thermodynamic derivative of the Ledoux criterion $\left(\partial P/ \partial \ye\right)_{P,s}$ is predominantly positive throughout the PCS convection zone when using the CMF EoS. This results in a straightforward classification of the electron fraction as acting either destabilising ($\D\ye/\D r > 0$) or stabilising ($\D\ye/\D r < 0$). The negative electron fraction gradient established at the outer convective boundary acts as \textit{destabilising}; the positive gradient toward the center of the PCS convection zone acts as \textit{stabilising}. Throughout the convective region, the radial entropy gradient $\D s/\D r < 0$, counteracting the stabilising $\ye$ gradient towards the center of the PCS convection zone. 
Additionally, we observe traces of LESA, the dipolar asymmetry in the electron fraction. Interestingly, we find Schwarzschild convection in this region, characterised by plumes with low entropy (destabilising) and high electron fraction (stabilising) sinking inward. This suggests that LESA may represent a unique manifestation of convection, potentially linked to the anomalous mixing behaviour of passive scalars.

In the following section, we compared self-consistent Favre-averaged 3D fluxes with the predictions of MLT and found that large-scale eddies are inadequately represented in regions where the local gradients of electron fraction $\D\ye/\D r$ and entropy $\D s/\D r$ are approximately flat. A relatively large dispersion characterizes these regions, but those exhibit relatively shallow slopes due to the Favre averaging used for the radial gradients. As a result, large eddies transporting material with high or low electron fractions into these regions are not well-captured by the MLT approach. We discussed that the dispersion $\delta\ye$ serves as a measure of the average eddy size and the distance it travels. In contrast, the conventional mixing length parameter remains constant and does not account for varying length scales. We link the larger eddy sizes 
with eddies travelling from afar with little mixing along the way, causing a higher dispersion, as observed towards the center of the PCS convection zone. Consequently, we suggest that a scale-dependent mixing length treatment might capture those larger eddies, travelling beyond one mixing length, more accurately. 

By decomposing the electron fraction into spherical harmonics, we find a shallower slope in their power spectrum, particularly at intermediate polynomial degrees ($6\lesssim l \lesssim 13$), compared to the velocity spectrum. We attribute this anomalous scaling to the (approximate) passive scalar nature of the electron fraction, which exhibits non-Kolmogorov typical scaling behaviour with a $l^{-1}$ slope rather than the expected $l^{-5/3}$ scaling law \citep{batchelor_1959}. The mismatched scaling between the velocity field and electron fraction can be explained by assuming that the electron contrast scales with eddy size but is independent of the underlying velocity field. We found that a symmetric eddy with a wavelength $\lambda$ scales with the mixing length parameter $\Lambda$ as $1/l$ when $\lambda\approx \Lambda$. Consequently, we concluded that a scale-dependent mixing length parameter could better account for the observed $l^{-1}$ scaling, which corresponds to larger eddies transporting material with high or low electron fractions over longer distances with minimal mixing along their way. This behaviour aligns with our observations and known characteristics of passive scalars \citep{Shraiman_1994}. 

\section*{Acknowledgements}
We acknowledge fruitful discussions with Rosemary Mardling. PJ is funded by the European Research
Council (ERC) Advanced Grant INSPIRATION under
the European Union’s Horizon 2020 research and innovation programm (Grant agreement No. 101053985).
BM was supported by ARC Future Fellowship FT160100035.  We acknowledge computer time allocations from Astronomy Australia Limited's ASTAC scheme, the National Computational Merit Allocation Scheme (NCMAS), and
from an Australasian Leadership Computing Grant.
Some of this work was performed on the Gadi supercomputer with the assistance of resources and services from the National Computational Infrastructure (NCI), which is supported by the Australian Government, and through support by an Australasian Leadership Computing Grant.  Some of this work was performed on the OzSTAR national facility at Swinburne University of Technology.  OzSTAR is funded by Swinburne University of Technology and the National Collaborative Research Infrastructure Strategy (NCRIS).

\section*{Data Availability}

The data from our simulations will be made available upon reasonable requests made to the authors.

\bibliographystyle{mnras}
\bibliography{example}

\appendix

\section{Approximation for the lapse function $\alpha$ at the edge of the PCS core}\label{appendix:alpha}
In order to estimate the lapse function $\alpha$
at the edge of the PCS core, we assume the weak field approximation in which $\alpha^2 \approx 1 - 2\Phi$, where $\Phi$ is the Newtonian gravitational potential at the location of the core $g$-mode. We then
approximately evaluate the Newtonian gravitational potential using the Poisson equation.
To compute $\Phi$ we need to account for PCS mass located both \emph{inside} and \emph{outside} the region where the $g$-mode lives, i.e., $M_\mathrm{mode} {\leq} M {\leq} M_\mathrm{PCS}$, as the density profile in the outer region will have a non-negligible contribution on the gravitational potential.  

The general solution for the Poisson equation can be written in the Green's function formalism as
\begin{align}\label{eq:gravpot}
    \Phi(\bm r) = -G\int\mathcal{G}(\bm r- \bm r')\rho(\bm r') \D \bm r'^3,
\end{align}
where $\mathcal{G}(\bm r-\bm r') = 1/|\bm r-\bm r'|$ satisfies $\Delta  \Phi(\bm r) = 4\pi \rho(\bm r') \delta(\bm r - \bm r')$. The approximate solution of the gravitational potential is calculated by decomposing the corresponding Green's function $\mathcal{G}$ into spherical harmonics analogously to \citet{steinmetz_1995}
\begin{align}
    \frac{1}{|\bm{r}-\bm{r}'|}=\sum_{l=0}^\infty \sum_{m=-l}^l\frac{4\pi}{2l+1}\mathcal{Y}_{lm}(\theta,\phi)\mathcal{Y}^*_{lm}(\theta'\phi')\frac{\min{(r,r')}}{\max{(r,r')}}.
\end{align}
The general solution of the Poisson equation is given by
\begin{align}
\nonumber
    \Phi(r,\theta,\phi) &= - G \sum_{l=0}^\infty \sum_{m=-l}^l\frac{4\pi}{2l+1} \mathcal{Y}_{lm}(\theta,\phi)\int_{4\pi}\int_0^\infty \D r' \D \Omega' \\
    &\times\rho (r',\theta,\phi)r'^2\rho(r')\frac{\min{(r,r')}^l}{\max{(r,r')}^{l+1}}\mathcal{Y}_{lm}^*(\theta,\phi),\label{eq:poisson}\\
    \intertext{where $\D\Omega = \sin\theta\,\D\theta\,\D\phi$ and $\min(r,r')^l = r^l$ for $r'\in [0,r]$ and $\max(r,r')^{-l-1} = r^{-l-1}$ for $r'\in [r,\infty]$. This is equivalent to  splitting the integral into an interior and exterior solution such as}
    \Phi(r,\theta,\phi) &= - G \sum_{l=0}^\infty \sum_{m=-l}^l\frac{4\pi}{2l+1} \mathcal{Y}_{lm}(\theta,\phi)\left(\frac{1}{r^{l+1}} C^{lm}(r) + r^l D^{lm}(r)\right), 
\end{align}
where the coefficients $C^{lm}$ and $D^{lm}$ only depend on radius,
\begin{align}
     C^{lm}(r) &= \int_{4\pi}\D\Omega'\mathcal{Y}^*_{lm}(\theta',\phi')\int_0^r\D r' r'^{l+2}\rho(r',\theta',\phi') \\ 
     D^{lm}(r) &= \int_{4\pi}\D\Omega'\mathcal{Y}^*_{lm}(\theta',\phi')\int_r^\infty\D r' r'^{1-l}\rho(r',\theta',\phi'). 
\end{align}
We now discard all higher order terms $l>0$ and omit any contributions to the potential by mass that is located outside of the neutron star with radius $R_\mathrm{NS}$ which gives 
\begin{align}
    \Phi_\mathrm{eff}(r) &\approx -4\pi G\left(\frac{1}{r_\mathrm{mode}}\int_0^{R_\mathrm{mode}} \D r' r'^2 \rho(r') + G\int_{R_\mathrm{mode}}^{R_\mathrm{NS}} \D r' r'\rho(r')\right) \nonumber \\ 
    &\approx -\frac{G M_\mathrm{mode}}{R_\mathrm{mode}} - \frac{4\pi G}{\langle r\rangle}\int_{R_\mathrm{mode}}^{R_\mathrm{NS}} \D r' {r'^2} \rho(r')
    {\langle r \rangle} \nonumber\\
    &= -\frac{G M_\mathrm{mode}}{R_\mathrm{mode}} - \frac{G(M_\mathrm{NS}-M_\mathrm{mode})}{\langle r\rangle},
\end{align}
where $\mathcal{Y}_{00} = 1/(2\sqrt{\pi})$. The quantity $\langle r\rangle$ represents an ``effective'' radius
for the material that is located between the inner $g$-mode and the neutron star surface,
\begin{equation}
\langle r\rangle=
    \frac{
    \int_{R_\mathrm{mode}}^{R_\mathrm{NS}} \D r' {r'}^2 \rho(r')}
    {\int_{R_\mathrm{mode}}^{R_\mathrm{NS}} \D r' {r'} \rho(r')}
\end{equation}
We can expect $\langle r\rangle $ to scale with
$R_\mathrm{mode}$ and
$R_\mathrm{NS}$, and parameterise $\langle r \rangle$ as 
\begin{align}
    \langle r\rangle \equiv c_\mathrm{eff} (R_\mathrm{NS} + R_\mathrm{mode}),
\end{align}
with the coefficient $c_\mathrm{eff} = 0.5$. Putting everything together, we read for our approximate lapse function 
\begin{align}
    \alpha &\approx \left[1 - 2G\left(\frac{M_\mathrm{mode}}{r_\mathrm{mode}} - \frac{M_\mathrm{NS} - M_\mathrm{mode}}{\langle r\rangle}\right)\right]^{1/2} \\
    &= 1 - G\left(\frac{M_\mathrm{mode}}{r_\mathrm{mode}} - \frac{M_\mathrm{NS} - M_\mathrm{mode}}{\langle r\rangle}\right) + \mathcal{O}(r^{-2}).
    \nonumber
\end{align}

\section{Condition for stabilising $\ye$ gradient from Maxwell relations}\label{subsec:mr}
The condition for instability in the Ledoux criterion is given by 
\begin{align}
    C_\mathrm{L} = \left(\pdv{\rho}{s}\right)_{\ye,P}\dv{s}{r} + \left(\pdv{\rho}{\ye}\right)_{s,P}\dv{\ye}{r} > 0.\label{eq:original}
\end{align}
We want to rewrite this criterion for instability in terms of the chemical potential $\mu_\nu$.
To this end, we can reformulate the thermodynamic derivative in the second term    
\begin{align}
\left(\pdv{\rho}{\ye}\right)_{P,s} &= -\frac{\partial(\rho,P,s)}{\partial(\ye,\rho,s)}\frac{\partial(\rho,\ye,s)}{\partial(\ye,P,s)}
%\\&
= -\frac{\partial(P,\rho,s)}{\partial(\ye,\rho,s)}\frac{\partial(\rho,\ye,s)}{\partial(P,\ye,s)} \nonumber \\
&= -\underbrace{\left(\pdv{P}{\ye}\right)_{\rho,s}}_{=\rho^2\left(\partial\mu_\nu/\partial\rho\right)_{\ye,s}}\left(\pdv{\rho}{P}\right)_{\ye,s} 
%\\ &
= -\rho^2\left(\pdv{\mu_\nu}{P}\right)_{\ye,s},\label{eq:pmu}
\end{align}
where we used the Maxwell relation $\left(\partial P/\partial \ye\right)_{\rho,s}=\rho^2\left(\partial\mu_\nu/\partial\rho\right)_{\ye,s}$. Substituting Equation~(\ref{eq:pmu}) into Equation~(\ref{eq:original}) yields
\begin{equation}
   C_\mathrm{L} = \left(\pdv{\rho}{s}\right)_{\ye,P}\dv{s}{r} -\rho^2\left(\pdv{\mu_\nu}{P}\right)_{\ye,s}\dv{\ye}{r}.\label{eq:reformed}
\end{equation}
Using the cyclic relation 
\begin{align}
    \left(\pdv{P}{s}\right)_{\ye,\rho}=-\left(\pdv{P}{\rho}\right)_{\ye,s} \left(\pdv{s}{P}\right)_{\ye,P}, 
\end{align} 
we can multiply Equation~(\ref{eq:reformed}) with the squared adiabatic sound speed $c_s^2=\left(\partial{P}/{\partial\rho}\right)_{\ye,s}$ (which is strictly positive) to obtain for the onset of Ledoux stability
\begin{align}
    C_\mathrm{L} =&-\underbrace{\left(\pdv{P}{\rho}\right)_{\ye,\rho} \left(\pdv{\rho}{s}\right)_{\ye,P}}_{-\left(\partial P/\partial s\right)_{\ye,\rho}}\dv{s}{r} -\rho^2\left(\pdv{P}{\rho}\right)_{\ye,s}\left(\pdv{\mu_\nu}{P}\right)_{\ye,s}\dv{\ye}{r} >0 \nonumber \\
    =& \left(\pdv{P}{s}\right)_{\ye,\rho}\dv{s}{r} - \rho^2\left(\pdv{\mu_\nu}{\rho}\right)_{\ye,s}>0.
\end{align}

\bsp	
\label{lastpage}
\end{document}